\documentclass[11pt]{article}
\usepackage{amsmath}
\usepackage{amssymb}
\usepackage{amsfonts}
\usepackage{graphics}
\usepackage{graphicx}
\usepackage{caption} 

\begin{document}

\title{$m$-step rational extensions of the trigonometric Darboux-P\"{o}%
schl-Teller potential based on para-Jacobi polynomials}
\author{Y.\ Grandati{\small \textsl{$^{1}$}} and C. Quesne{\small \textsl{$%
^{2}$}} \\
{\small \textsl{$^{1}$LPCT UMR 7019, Universit\'{e} de
Lorraine Site de Metz,}}\\
{\small \textsl{1 bvd D. F. Arago, F-57070, Metz, France}}\\
{\small \textsl{yves.grandati@univ-lorraine.fr}}\\
{\small \textsl{$^{2}$D\'{e}partement de Physique, Universit\'{e} Libre de
Bruxelles,}} \\
{\small \textsl{Campus de la Plaine CP229, Boulevard~du Triomphe, B-1050
Brussels, Belgium}}\\
{\small \textsl{christiane.quesne@ulb.be}}}
\date{ }
\maketitle

\begin{abstract}
A previous construction of regular rational extensions of the trigonometric
Darboux-P\"oschl-Teller potential, obtained by one-step Darboux
transformations using seed functions associated with the para-Jacobi
polynomials of Calogero and Yi, is generalized by considering $m$-step
Darboux transformations. As a result, some novel families of exceptional
orthogonal polynomials depending on $m$ discrete parameters, as well as $m$
continuous real ones $\lambda_1$, $\lambda_2$, \ldots, $\lambda_m$, are
obtained. The restrictions imposed on these parameters by the rational
extensions regularity conditions are studied in detail.
\end{abstract}

\baselineskip=22pt plus 1pt minus 1pt 

\baselineskip=22pt plus 1pt minus 1pt 
\newpage

\section{INTRODUCTION}

Exceptional orthogonal polynomials (EOPs) are complete families of
orthogonal polynomials that arise as eigenfunctions of a Sturm-Liouville
eigenvalue problem \cite{gomez09}. Their most apparent difference with
respect to the classical orthogonal polynomials (COPs) of Jacobi, Laguerre,
and Hermite is that there are some gaps in the sequence of their degrees,
the total number of missing ``exceptional'' degrees being known as the
codimension. In addition, the corresponding differential equation contains
rational coefficients instead of polynomial ones.

%
%
Since their appearance, EOPs have generated a lot of research activity both
from a mathematical viewpoint and for their applications in mathematical
physics. In mathematics, studies have been carried out on the properties of
their zeros \cite{gomez13a, kuijlaars}, the sets of recurrence relations
they satisfy \cite{duran15, miki, odake16}, and the search for a full
classification \cite{gomez13b, garcia19}. In mathematical physics, EOPs were
shown to be related to the Darboux transformation (DT) in the construction
of exactly solvable rationally-extended quantum potentials \cite{cq08, cq09,
odake09}. This led to multi-indexed families of EOPs connected with
multi-step DT \cite{gomez12, odake11, cq11, grandati12}. EOPs also appeared
in connection with exact solutions of the Dirac equation \cite{schulze},
some superintegrable systems \cite{marquette, post}, or rational solutions
of some Painlev\'e equations \cite{clarkson}.

%
%
Before 2015, it was thought that apart from the parameters associated with
the corresponding COPs, there only appeared discrete parameters in the
construction of EOPs. In the Jacobi family case, for instance, the most
general known construction was given in terms of a Wronskian-like
determinant of classical Jacobi polynomials, indexed by two partitions \cite%
{bonneux, duran17}. Some possibilities of having more general EOP families,
also depending on continuous parameters, then made their appearance.

%
%
First, on considering the trigonometric Darboux-P\"oschl-Teller (TDPT)
potential, whose bound states are expressed in terms of Jacobi polynomials,
and using the fact, first noticed by Szeg\"o \cite{szego} and emphasized by
Calogero and Yi \cite{calogero}, that for some values of the parameters the
Jacobi equation has general polynomial solutions depending on an arbitrary
real parameter $\lambda$ and called para-Jacobi polynomials, it was possible
to build some new regular rational extensions of the TDPT potential by
one-step DT \cite{bagchi}. The eigenstates of such extensions were shown to
be associated with a novel family of $\lambda$-dependent EOPs.

%
%
Second, in the cases of the TDPT and isotonic potentials, some new rational
extensions were constructed by considering two-step confluent chains of DT,
that is chains of DT in which the spectral parameters of the different seed
functions converge to the same value \cite{grandati15}. Both types of
extended potentials are associated with new families of orthogonal
polynomials. In the case of the TDPT potential, the latter depend on a
continuous real parameter. Later on, the confluent DT algorithm was also
used to construct exceptional Legendre polynomials \cite{garcia21}, as well
as exceptional Gegenbauer ones \cite{garcia22} with an arbitrary number of
real parameters.

%
%
Third, a dualization of Krall dual Hahn polynomials was used to construct
exceptional Hahn polynomials, which in the limit led to exceptional Jacobi
polynomials depending on an arbitrary number of continuous parameters \cite%
{duran21}. The latter were then shown \cite{duran23} to be divided in two
classes: some families that are deformations of Jacobi polynomials, as those
determined in \cite{grandati15}, and other families that are deformations of
standard families of Jacobi EOPs.

%
%
The purpose of the present paper is to complete the study made in Ref.~\cite%
{bagchi}, which was limited to one-step DT. Here, we plan to study in detail
its multi-step version, which will allow us to derive novel families of
exceptional para-Jacobi polynomials depending on an arbitrary number of free
continuous parameters.

%
%
The paper is organized as follows. We start by recalling the essential
features of DT and of the TDPT potential in Secs.~II and III, respectively.
In Sec.~IV, we summarize the results obtained for one-step regular
extensions of the TDPT potential based on para-Jacobi polynomials. In
Sec.~V, we study in detail two-step regular extensions of the same. In
Sec.~VI, a similar analysis is carried out for $m$-step regular extensions.
Finally, Sec.~VII contains the conclusion.

%
%

\section{DARBOUX TRANSFORMATIONS (DT)}

\subsection{One-step DT}

We consider a one-dimensional Hamiltonian $\widehat{H}=-d^{2}/dx^{2}+V(x)$, $%
x\in I\subset \mathbb{R}$ and the associated Schr\"{o}dinger equation 
\begin{equation}
\psi _{\lambda }^{\prime \prime }(x)+(E_{\lambda }-V(x))\psi _{\lambda
}(x)=0,
\end{equation}
where $\psi _{\lambda }(x)$ is a formal eigenfunction of $\widehat{H}$ for
the eigenvalue $E_{\lambda }$ and is defined up to a multiplicative
constant. In the following, we suppose that with Dirichlet boundary
conditions on $I$, $\widehat{H}$ admits a discrete spectrum of energies and
eigenstates $(E_{n};\psi _{n})_{n\in \left\{ 0,...,n_{\max }\right\} \subset 
\mathbb{N}}$ where, without loss of generality, we can always suppose that
the ground level of $\widehat{H}$ is zero: $E_{0}=0$.

%
%
Any formal eigenfunction $\psi _{\nu }(x)$ of $V(x)$(i.e., of $\widehat{H}$)
can be used as a seed function for a Darboux transformation (DT) $A(\psi
_{\nu })$, which associates with the potential $V(x)$ a modified potential 
\begin{equation}
V(x)\overset{A(\psi _{\nu })}{\longrightarrow }V^{(\nu )}(x)=V(x)-2(\log
(\psi _{\nu }(x)))^{\prime \prime },
\end{equation}%
which we call an extension of $V(x)$. The formal eigenfunction of $V^{(\nu
)} $ associated with the spectral parameter $E_{\lambda }$ is given by the
Darboux-Crum formulas \cite{crum} 
\begin{equation}
\left\{ 
\begin{array}{c}
\psi _{\lambda }^{\left( \nu \right) }(x)\sim \frac{W\left( \psi _{\nu
},\psi _{\lambda }\mid x\right) }{\psi _{\nu }(x)},\text{ if }\lambda \neq
\nu , \\ 
\psi _{\nu }^{\left( \nu \right) }(x)\sim \frac{1}{\psi _{\nu }(x)},%
\end{array}%
\right.
\end{equation}%
where $W(y_{1},...,y_{m}\mid x)$ denotes the Wronskian of the family of
functions $y_{1}(x),...,y_{m}(x)$ \cite{muir}. 
\begin{equation}
W\left( y_{1},...,y_{m}\mid x\right) =\left\vert 
\begin{array}{ccc}
y_{1}\left( x\right) & ... & y_{m}\left( x\right) \\ 
... &  & ... \\ 
y_{1}^{\left( m-1\right) }\left( x\right) & ... & y_{m}^{\left( m-1\right)
}\left( x\right)%
\end{array}%
\right\vert .
\end{equation}

This last possesses the following useful properties \cite{muir}

\begin{equation}
\left\{ 
\begin{array}{c}
W\left( uy_{1},...,uy_{m}\mid x\right) =u(x)^{m}W\left( y_{1},...,y_{m}\mid
x\right) \\ 
W\left( y_{1},...,y_{m}\mid x\right) =\left( \frac{dz}{dx}\right)
^{m(m-1)/2}W\left( y_{1},...,y_{m}\mid z\right) .%
\end{array}%
\right.  \label{wronsk prop}
\end{equation}

%
%

\subsection{Chains of DT}

At the formal level, the DT can be straightforwardly iterated and a chain of 
$m$ DT can be completely characterized by the $m$-tuple $(\nu _{1},...,\nu
_{m})$ of spectral indices of the successive seed functions used in the
chain. We denote such a $m$-tuple by a capital letter $N_{m}$, where the
index $m$ indicates the length of the chain $N_{m}$. Let $\psi _{\lambda
}^{\left( N_{m}\right) }$ be the formal eigenfunction associated with the
eigenvalue $E_{\lambda }$ of the potential $V^{\left( N_{m}\right) }(x)$.

%
%
A chain is non-degenerate if all the spectral indices $\nu _{i}$ of the
chain $N_{m}$ are distinct and is degenerate if some of them are repeated in
the chain. For non-degenerate chains, Crum has derived very useful formulas
for the extended potentials and their eigenfunctions in terms of Wronskians
of eigenfunctions of the initial potential \cite{crum}.

\textbf{Crum's formulas}

\textit{When all the }$\nu _{j}$\textit{\ and }$\lambda $\textit{\ are
distinct, we have} 
\begin{equation}
\psi _{\lambda }^{\left( N_{m}\right) }(x)=\frac{W^{\left( N_{m},\lambda
\right) }\left( x\right) }{W^{\left( N_{m}\right) }\left( x\right) }
\label{etats n3}
\end{equation}
\textit{and} 
\begin{equation}
V^{\left( N_{m}\right) }(x)=V(x)-2\left( \log W^{\left( N_{m}\right) }\left(
x\right) \right) ^{\prime \prime },  \label{potnstep2}
\end{equation}
\textit{where }$W^{\left( N_{m}\right) }\left( x\right) =W\left( \psi _{\nu
_{1}},...,\psi _{\nu _{m}}\mid x\right) $\textit{.}

%
%
These formulas can be extended to degenerate chains if we adopt the
convention to suppress any pair of repeated indices in the set of spectral
indices associated with the chain.

%
%

\section{TDPT POTENTIAL}

\setcounter{equation}{0}

\subsection{Spectrum}

The trigonometric Darboux-P\"{o}schl-Teller (TDPT) potential (with zero
ground-state energy) is defined on $]0,\pi /2[$ by 
\begin{equation}
V(x;\alpha ,\beta )=\frac{\left( \alpha +1/2\right) (\alpha -1/2)}{\sin
^{2}x }+\frac{\left( \beta +1/2\right) (\beta -1/2)}{\cos ^{2}x}-(\alpha
+\beta +1)^{2},
\end{equation}
and is a confining potential for $\left\vert \alpha \right\vert ,\left\vert
\beta \right\vert >1/2$. On introducing the variable $z=\cos 2x$, it can be
rewritten as 
\begin{equation}
V(x;\alpha ,\beta )=\frac{2\left( \alpha +1/2\right) (\alpha -1/2)}{1-z}+ 
\frac{2\left( \beta +1/2\right) (\beta -1/2)}{1+z}-(\alpha +\beta +1)^{2}.
\end{equation}

%
%
{}For $\alpha ,\beta >1/2$, the physical spectrum of the TDPT potential
associated with the asymptotic Dirichlet boundary conditions 
\begin{equation}
\psi \left( 0^{+};\alpha ,\beta \right) =0=\psi \left( \left( \pi /2\right)
^{-};\alpha ,\beta \right)  \label{eq:dirichlet}
\end{equation}
is given in terms of Jacobi polynomials $P_{n}^{\left( \alpha ,\beta \right)
}(z)$ by $\left( \left( \alpha _{n},\beta _{n}\right) =\left( \alpha
+n,\beta +n\right) \right) $ 
\begin{equation}
\left\{ 
\begin{array}{c}
\psi _{n}\left( x;\alpha ,\beta \right) =\psi _{0}\left( x;\alpha ,\beta
\right) P_{n}^{\left( \alpha ,\beta \right) }\left( z\right) \\ 
E_{n}\left( \alpha ,\beta \right) =\left( \alpha _{n}+\beta _{n}+1\right)
^{2}-\left( \alpha +\beta +1\right) ^{2}=4n\left( \alpha +\beta +1+n\right)%
\end{array}
\right. ,n\in \mathbb{N},
\end{equation}
where 
\begin{equation}
\psi _{0}\left( x;\alpha ,\beta \right) =2^{\alpha +\beta +1}\left( \sin
x\right) ^{\alpha +1/2}\left( \cos x\right) ^{\beta +1/2}=(1-z)^{\left(
\alpha +1/2\right) /2}(1+z)^{\left( \beta +1/2\right) /2}
\end{equation}
and 
\begin{eqnarray}
P_{n}^{\left( \alpha ,\beta \right) }\left( z\right) &=&\frac{1}{2^{n}}
\sum_{k=0}^{n}\left( -1\right) ^{n-k}\binom{n+\alpha }{k}\binom{ n+\beta }{%
n-k}(1-z)^{n-k}(1+z)^{k}  \notag \\
&=&\frac{\left( -1\right) ^{n}\Gamma \left( n+\beta +1\right) }{n!\Gamma
\left( n+\alpha +\beta +1\right) }\sum_{k=0}^{n}\left( -1\right) ^{k} \binom{%
n}{k}\frac{\Gamma \left( n+\alpha +\beta +1+k\right) }{2^{k}\Gamma \left(
\beta +1+k\right) }(1+z)^{k}.
\end{eqnarray}
The quasipolynomial eigenfunction (i.e., which is, up to a gauge factor,
polynomial in the adapted variable $z$) $\psi _{n}\left( x;\alpha ,\beta
\right) $ is then the particular solution of the Schr\"{o}dinger equation 
\begin{equation}
\left( -\frac{d^{2}}{dx^{2}}+V\left( x;\alpha ,\beta \right) -E_{n}\left(
\alpha ,\beta \right) \right) \psi \left( x\right) =0,
\end{equation}
which satisfies the Dirichlet boundary conditions (\ref{eq:dirichlet}).

%
%

\subsection{Para-Jacobi polynomials}

Suppose that $\alpha $ and $\beta $ are two positive integers 
\begin{equation}
\alpha =N\in \mathbb{N}^{\ast },\qquad \beta =M\in \mathbb{N}^{\ast }.
\end{equation}
In this case, as noticed by Szeg\"{o} \cite{szego} and emphasized by
Calogero and Yi \cite{calogero, bagchi}, for values of $n$ such that \cite{footnote} 
\begin{equation}
  \max \left( N,M\right) \leq n<N+M,
\end{equation}
one obtains the general solution of the Schr\"{o}dinger equation 
\begin{equation}
  \left( -\frac{d^{2}}{dx^{2}}+V\left( x;N,M\right) -E_{-n-1}\left( N,M\right)
  \right) \psi \left( x\right) =0,
\end{equation}
which is of the quasipolynomial form 
\begin{equation}
  \psi _{-n-1}\left( z;N,M;\lambda \right) =\psi _{-1}\left( z;N,M\right)
  p_{n}^{\left( -N,-M\right) }\left( z;\lambda \right) ,  \label{eq:general}
\end{equation}
where, up to a constant factor, 
\begin{align}
  \psi _{-1}\left( z;N,M\right) &=\psi _{0}\left( z;-N,-M\right) \nonumber\\
 &=(1-z)^{\left(-N+1/2\right) /2}(1+z)^{\left( -M+1/2\right) /2} \nonumber\\
 &=\frac{1}{\psi _{0}\left(z;N-1,M-1\right) }.  \label{psi -1}
\end{align}
\par
%
%
In Eq.~(\ref{eq:general}), $p_{n}^{\left( -N,-M\right) }\left( z;\lambda
\right) $ is a polynomial depending on an arbitrary real parameter $\lambda $
and is called \textbf{para-Jacobi polynomial} (In the following, $\psi
_{-n-1}$ will be called a \textbf{para-Jacobi (PJ) eigenfunction}). It has
the (monic) form \cite{calogero} 
\begin{eqnarray}
p_{n}^{\left( -N,-M\right) }\left( z;\lambda \right) &=&\frac{\left(
-2\right) ^{n}\left( n-M\right) !n!}{\left( 2n-M-N\right) !}\Theta
_{n,1}^{\left( -N,-M\right) }(z)  \notag \\
&&+\lambda \frac{\left( -2\right) ^{n}\left( 2n-M-N+1\right) !\left(
M+N-n-1\right) !}{\left( n-N\right) !}\Theta _{n,2}^{\left( -N,-M\right)
}(z),  \label{eq:para-Jacobi}
\end{eqnarray}
where 
\begin{equation}
\left\{ 
\begin{array}{c}
\Theta _{n,1}^{\left( -N,-M\right) }(z)=\sum_{k=M}^{n}\frac{\left(
-1\right)^{k}\left( n-M-N+k\right) !}{2^{k}k!\left(k-M\right)! \left(
n-k\right) !}(1+z)^{k} \\ 
\Theta _{n,2}^{\left( -N,-M\right) }(z)=\sum_{k=0}^{N+M-n-1}\frac{ \left(
-1\right) ^{k}\left( M-1-k\right) !}{2^{k}k!\left( N+M-n-1-k\right)! \left(
n-k\right) !}(1+z)^{k}%
\end{array}
\right. ,
\end{equation}
with $N,M>0$, and 
\begin{equation}
\left\{ 
\begin{array}{c}
\Theta _{n,1}^{\left( -N,-M\right) }(-1)=0 \\ 
\Theta _{n,2}^{\left( -N,-M\right) }(-1)=\frac{\left( M-1\right) !}{\left(
N+M-n-1\right)! n!}%
\end{array}
\right. .
\end{equation}
\par
%
%
We have in particular 
\begin{equation}
p_{n}^{\left( -N,-M\right) }\left( -1;\lambda \right) =\left( -1\right)
^{n}\lambda b_{n}^{\left( N,M\right) },  \label{eq:para-1}
\end{equation}
where 
\begin{equation}
b_{n}^{\left( N,M\right) }=\frac{2^{n}\left( 2n-N-M+1\right) !\left(
M-1\right) !}{n!\left( n-N\right) !}.  \label{eq:b}
\end{equation}

%
%
The para-Jacobi polynomials satisfy the derivation property 
\begin{equation}
\overset{.}{p}_{n}^{\left( -N,-M\right) }\left( z;\lambda \right)
=np_{n-1}^{\left( -N+1,-M+1\right) }\left( z;a_{n}^{\left( N,M\right)
}\lambda \right) ,  \label{eq:para-der}
\end{equation}%
with 
\begin{equation}
a_{n}^{\left( N,M\right) }=\frac{M+N-n-1}{n}.  \label{eq:a}
\end{equation}

%
%
We also have the symmetry property \cite{calogero} 
\begin{equation}
p_{n}^{\left( -N,-M\right) }\left( -z;\lambda \right) =\left( -1\right)
^{n}p_{n}^{\left( -M,-N\right) }\left( z;g_{n}^{\left( N,M\right) }\left(
\lambda \right) \right) ,  \label{eq:para-sym}
\end{equation}
where $g_{n}^{\left( N,M\right) }$ is the affine function 
\begin{equation}
g_{n}^{\left( N,M\right) }\left( \lambda \right) =\left( -1\right)
^{n-M}\left( \left( -1\right) ^{n-N+1}\lambda +\lambda _{n}^{\left(
N,M\right) }\right) ,  \label{eq:g}
\end{equation}
with 
\begin{equation}
\lambda _{n}^{\left( N,M\right) }=\frac{n!\left( n-M\right) !\left(
n-N\right) !}{\left( 2n-N-M\right) !\left( 2n-N-M+1\right) !\left(
N+M-n-1\right) !}>0.  \label{eq:lambda}
\end{equation}
This implies in particular (cf Eq.~(\ref{eq:para-1})) 
\begin{equation}
p_{n}^{\left( -N,-M\right) }\left( 1;\lambda \right) =b_{n}^{\left(
N,M\right) }g_{n}^{\left( N,M\right) }\left( \lambda \right) .
\label{eq:para1}
\end{equation}

%
%
Note the following useful identities 
\begin{equation}
\lambda _{n-1}^{\left( N-1,M-1\right) }=a_{n}^{\left( N,M\right) }\lambda
_{n}^{\left( N,M\right) },  \label{eq:lambda-lambda}
\end{equation}
\begin{equation}
b_{n-1}^{\left( N-1,M-1\right) }=\frac{n}{2\left( M-1\right) }b_{n}^{\left(
N,M\right) },  \label{eq:b-b}
\end{equation}
and 
\begin{equation}
b_{n}^{\left( M,N\right) }=\frac{\left( N-1\right) !\left( n-N\right) !}{
\left( M-1\right) !\left( n-M\right) !}b_{n}^{\left( N,M\right) }.
\end{equation}

%
%

\section{ONE-STEP REGULAR EXTENSIONS OF THE TDPT POTENTIAL}

\setcounter{equation}{0}

{}For $\max(N,M)\leq n<N+M$, we have $E_{-n-1}(N,M)<0$ and $\psi
_{-n-1}\left( z;N,M;\lambda \right) $ is disconjugated on $x\in $ $]0,\pi
/2[ $ $(z\in $ $]-1,+1[)$ and consequently admits at most one zero on this
interval. As proven in \cite{bagchi}, $\psi _{-n-1}$ has no node there iff
the parameters satisfy the following \textbf{regularity conditions}: 
\begin{equation}
\left\{ 
\begin{array}{c}
(M,n-N)\in \left( 2\mathbb{N}\right) ^{2}:0<\lambda <\lambda _{n}^{\left(
N,M\right) }; \\[0.2cm] 
(M,n-N)\in 2\mathbb{N}\times \left( 2\mathbb{N}+1\right) :\lambda <-\lambda
_{n}^{\left( N,M\right) }\text{ or }\lambda >0; \\[0.2cm] 
(M,n-N)\in \left( 2\mathbb{N}+1\right) \times 2\mathbb{N}:\lambda <0\text{
or }\lambda >\lambda _{n}^{\left( N,M\right) }; \\[0.2cm] 
(M,n-N)\in \left( 2\mathbb{N}+1\right) ^{2}:-\lambda _{n}^{\left( N,M\right)
}<\lambda <0.%
\end{array}%
\right.  \label{eq:reg-1}
\end{equation}

%
%
With the choices above, we can then use $\psi _{- n-1}$ as seed function of
a one-step DT to produce a rational extension of the TDPT potential, which
is perfectly regular on $]0,\pi /2[$. This extended potential is given by 
\cite{bagchi} 
\begin{equation}
V^{(-n-1)}\left( x;N,M;\lambda \right) =V\left( x;N-1,M-1\right) -4(N+M)-2%
\frac{d^{2}}{dx^{2}}\log (p_{n}^{\left( -N,-M\right) }\left( z;\lambda
\right) )
\end{equation}
and can be continuously modulated by varying the parameter $\lambda $ on the
adapted interval.

%
%
Its spectrum is given by $E_{k}\left( N,M\right) $, $k\in \left\{
-n-1,0,1,...,\right\} $, with the corresponding eigenstates:
\begin{equation}
\psi _{-n-1}^{\left( -n-1\right) }\left( x;N,M;\lambda \right) =\frac{1}{%
\psi _{-n-1}\left( x;N,M;\lambda \right) }=\psi _{0}\left( x;N-1,M-1\right) 
\frac{1}{p_{n}^{(-N,-M)}(z;\lambda )}  \label{psi -n-1}
\end{equation}
and
\begin{eqnarray}
  && \psi _{k}^{\left( -n-1\right) }\left( x;N,M;\lambda \right) \nonumber \\
  && =\frac{%
     W\left( \psi _{-n-1}(x;N,M;\lambda ),\psi _{k}(x;N,M)\mid x\right) }{\psi
     _{-n-1}\left( x;N,M;\lambda \right) }  \nonumber \\
  && =\psi _{-1}(x;N,M)\frac{dz}{dx}\frac{W\left( p_{n}^{(-N,-M)}(z;\lambda
     ),(1-z)^{N}(1+z)^{M}P_{k}^{(N,M)}(z)\mid x\right) }{p_{n}^{(-N,-M)}(z;%
     \lambda )} \nonumber\\
  && \ \propto\frac{\psi _{-1}(x;N-1,M-1)}{p_{n}^{(-N,-M)}(z;\lambda )} \nonumber\\
  &&\times \left\vert 
    \begin{array}{cc}
    p_{n}^{(-N,-M)}(z;\lambda ) & (1-z)^{N}(1+z)^{M}P_{k}^{(N,M)}(z) \\ 
    np_{n-1}^{\left( -N+1,-M+1\right) }\left( z;a_{n}^{(N,M)}\lambda \right)  & 
    -2(k+1)(1-z)^{N-1}(1+z)^{M-1}P_{k+1}^{(N-1,M-1)}(z)%
   \end{array}%
   \right\vert
\end{eqnarray}
for $k=0, 1, \ldots$. Here we have used Eq.~(\ref{eq:para-der}), Eq.~(\ref{wronsk prop}), and the
following derivation rule of Jacobi polynomials \cite{koekoek} 
\begin{equation}
  \frac{d}{dz}\left( (1-z)^{N}(1+z)^{M}P_{k}^{(N,M)}(z)\right)
  =-2(k+1)(1-z)^{N-1}(1+z)^{M-1}P_{k+1}^{(N-1,M-1)}(z),  \label{deriv Jac 1}
\end{equation}
which also makes its appearance in connection with the shape
invariance of the TDPT potential\cite{GGM}.\par
%
%
Then
\begin{equation}
  \psi _{k}^{\left( -n-1\right) }\left( x;N,M;\lambda \right) =\frac{\psi
  _{0}(x;N-1,M-1)}{p_{n}^{(-N,-M)}(z;\lambda )}Q_{k}^{\left( n\right) }\left(
  z;N,M;\lambda \right) ,  \label{psi1-k}
\end{equation}
where the $Q_{k}^{\left( n\right) }$'s are given by $Q_{-n-1}^{\left(
n\right) }\left( z;N,M;\lambda \right) =1$ and, for $k\geq 0$,
\begin{eqnarray}
  Q_{k}^{\left( n\right) }\left( z;N,M;\lambda \right)  &=&\left\vert 
    \begin{array}{cc}
      p_{n}^{(-N,-M)}(z;\lambda ) & (1-z^{2})P_{k}^{(N,M)}(z) \\ 
      np_{n-1}^{\left( -N+1,-M+1\right) }\left( z;a_{n}^{(N,M)}\lambda \right)  & 
      -2(k+1)P_{k+1}^{(N-1,M-1)}(z)%
    \end{array}%
    \right\vert   \label{Q1} \\
  &=&-2(k+1)P_{k+1}^{(N-1,M-1)}(z)p_{n}^{(-N,-M)}(z;\lambda )  \notag \\
  &&-n(1-z^{2})P_{k}^{(N,M)}(z)p_{n-1}^{\left( -N+1,-M+1\right) }\left( z;%
\frac{M+N-n-1}{n}\lambda \right) ,
\end{eqnarray}
\par
%
%
The set of $Q_k^{(n)}(z;N,M;\lambda)$, $k=-n-1, 0, 1, 2, \ldots$, is made of
orthogonal polynomials on $]-1,1[$ with respect to the measure 
\begin{equation}
\mu _{n}^{\left( -N,-M\right) }\left( z;\lambda \right) =\frac{
(1-z)^{N-1}(1+z)^{M-1}}{\left( p_{n}^{\left( -N,-M\right) }\left( z;\lambda
\right) \right) ^{2}}.
\end{equation}
\par
%
%

\section{TWO-STEP REGULAR EXTENSIONS OF THE TDPT POTENTIAL}

\subsection{Regularity conditions}

\setcounter{equation}{0}

Consider now a two-step chain of DT based on PJ seed functions $\psi
_{-n_{1}-1 }\left( x;N,M;\lambda _{1}\right) $ and $\psi _{-n_{2}-1 }\left(
x;N,M;\lambda _{2}\right) $, 
\begin{equation}
V\left( x;N,M\right) \overset{A(\psi _{-n_{1}-1 })}{ \longrightarrow }%
V^{\left( -n_{1}-1\right) }\left( x;N,M;\lambda _{1}\right) \overset{%
A\left(\psi _{- n_{2}-1 }^{\left( -n_{1}-1\right) }\right)}{ \longrightarrow 
}V^{\left( -n_{1}-1,-n_{2}-1\right) }\left( x;N,M;\lambda _{1},\lambda
_{2}\right) ,
\end{equation}
where 
\begin{equation}
\psi _{-n_{2}-1 }^{\left( -n_{1}-1\right) }\left( x;N,M;\lambda _{1},\lambda
_{2}\right) =\frac{W\left( \psi _{- n_{1}-1}(x;N,M;\lambda_1),\psi
_{-n_{2}-1}(x;N,M;\lambda_2) \mid x\right) }{\psi _{- n_{1}-1}\left(
x;N,M;\lambda _{1}\right) }.
\end{equation}

%
%
We are interested in chains for which both extensions $V^{\left(
-n_{1}-1\right) }$ and $V^{\left( -n_{1}-1,-n_{2}-1\right) }$ are regular.
For such a purpose, we have first to choose $\max \left( N,M\right) \leq
n_{1}<N+M$ and $\lambda _{1}$ in order to satisfy the regularity condition (%
\ref{eq:reg-1}), which ensures that the seed function $\psi _{-n_{1}-1}$ has
no node on $]0,\pi /2[$. We then choose $n_{2}$ in order that the seed
function of the second DT $\psi _{-n_{2}-1}^{\left( -n_{1}-1\right) }$ be in
the disconjugacy sector of $V^{\left( -n_{1}-1\right) }$, namely 
\begin{equation}
E_{-n_{1}-1}\left( N,M\right) -E_{-n_{2}-1}\left( N,M\right)
=4(n_{1}-n_{2})\left( n_{1}+n_{2}-N-M+1\right) >0.  \label{eq:E-E}
\end{equation}%
This is achieved if 
\begin{equation}
\max \left( N,M\right) \leq n_{2}<n_{1}<N+M.  \label{eq:n_1-n_2}
\end{equation}

%
%
$V^{\left( -n_{1}-1,-n_{2}-1\right) }\left( x;N,M;\lambda _{1},\lambda
_{2}\right) $ is then regular when $\psi _{-n_{2}-1}^{\left( -n_{1}-1\right)
}\left( x;N,M;\lambda _{1},\lambda _{2}\right) $ is nodeless on $]0,\pi /2[$%
. Since $\psi _{-n_{1}-1 }$ does not change its sign on this interval, this
corresponds to 
\begin{equation}
\mathrm{sign}\left( W\left( \psi _{-n_{1}-1},\psi _{- n_{2}-1}\mid 0\right)
\right) =\mathrm{sign}\left( W\left( \psi _{- n_{1}-1},\psi _{- n_{2}-1}\mid
\pi /2\right) \right) .  \label{eq:reg-two-step}
\end{equation}
But we have 
\begin{eqnarray}
W\left( \psi _{-n_{1}-1},\psi _{-n_{2}-1}\mid x\right)
&=&-2(1-z)^{-N+1}(1+z)^{-M+1}  \notag \\
&&\times W\left( p_{n_{1}}^{\left( -N,-M\right) }\left( z;\lambda
_{1}\right) ,p_{n_{2}}^{\left( -N,-M\right) }\left( z;\lambda _{2}\right)
\mid z\right) ,
\end{eqnarray}
from which we deduce that the preceding condition can be rewritten as 
\begin{eqnarray}
&& \mathrm{sign}\left( W\left( p_{n_{1}}^{\left( -N,-M\right) }\left(
z;\lambda _{1}\right) ,p_{n_{2}}^{\left( -N,-M\right) }\left( z;\lambda
_{2}\right) \mid 1\right) \right)  \notag \\
&&=\mathrm{sign}\left( W\left( p_{n_{1}}^{\left( -N,-M\right) }\left(
z;\lambda _{1}\right) ,p_{n_{2}}^{\left( -N,-M\right) }\left( z;\lambda
_{2}\right) \mid -1\right) \right) .  \label{eq:reg-2}
\end{eqnarray}

%
%
On using Eqs.~(\ref{eq:para-1}), (\ref{eq:para-der}), (\ref{eq:a}), and (\ref%
{eq:b-b}), we obtain 
\begin{eqnarray}
&&W\left( p_{n_{1}}^{\left( -N,-M\right) }\left( z;\lambda _{1}\right)
,p_{n_{2}}^{\left( -N,-M\right) }\left( z;\lambda _{2}\right) \mid -1\right)
\notag \\
&=&\left\vert 
\begin{array}{cc}
\left( -1\right) ^{n_{1}}b_{n_{1}}^{\left( N,M\right) }\lambda _{1} & \left(
-1\right) ^{n_{2}}b_{n_{2}}^{\left( N,M\right) }\lambda _{2} \\ 
n_{1}\left( -1\right) ^{n_{1}-1}b_{n_{1}-1}^{\left( N-1,M-1\right)
}a_{n_{1}}^{\left( N,M\right) }\lambda _{1} & n_{2}\left( -1\right)
^{n_{2}-1}b_{n_{2}-1}^{\left( N-1,M-1\right) }a_{n_{2}}^{\left( N,M\right)
}\lambda _{2}%
\end{array}%
\right\vert  \notag \\
&=&\left( -1\right) ^{n_{1}+n_{2}}\frac{b_{n_{1}}^{\left( N,M\right)
}b_{n_{2}}^{\left( N,M\right) }}{2(M-1)}(n_{2}-n_{1})\left(
n_{1}+n_{2}-N-M+1\right) \lambda _{1}\lambda _{2},
\end{eqnarray}%
or, with Eq.~(\ref{eq:E-E}), 
\begin{eqnarray}
&&W\left( p_{n_{1}}^{\left( -N,-M\right) }\left( z;\lambda _{1}\right)
,p_{n_{2}}^{\left( -N,-M\right) }\left( z;\lambda _{2}\right) \mid -1\right)
\notag \\
&=&\left( -1\right) ^{n_{1}+n_{2}-1}\frac{b_{n_{1}}^{\left( N,M\right)
}b_{n_{2}}^{\left( N,M\right) }}{8(M-1)}\left( E_{-n_{1}-1}\left( N,M\right)
-E_{-n_{2}-1}\left( N,M\right) \right) \lambda _{1}\lambda _{2}.
\label{eq:W-1}
\end{eqnarray}

%
%
On the other hand (see Eqs.~(\ref{eq:para-1}), (\ref{eq:para-der}), (\ref%
{eq:para-sym}), and (\ref{eq:b-b})), $W\left( p_{n_{1}}^{\left( -N,-M\right)
}\left( z;\lambda _{1}\right) ,p_{n_{2}}^{\left( -N,-M\right) }\left(
z;\lambda _{2}\right) \mid +1\right) $ can be written as 
\begin{eqnarray}
&&W\left( p_{n_{1}}^{\left( -N,-M\right) }\left( z;\lambda _{1}\right)
,p_{n_{2}}^{\left( -N,-M\right) }\left( z;\lambda _{2}\right) \mid +1\right)
\notag \\
&=&\left\vert 
\begin{array}{cc}
p_{n_{1}}^{\left( -N,-M\right) }\left( 1;\lambda _{1}\right) & 
p_{n_{2}}^{\left( -N,-M\right) }\left( 1;\lambda _{2}\right) \\ 
n_{1}p_{n_{1}-1}^{\left( -N+1,-M+1\right) }\left( 1;a_{n_{1}}^{\left(
N,M\right) }\lambda _{1}\right) & n_{2}p_{n_{2}-1}^{\left( -N+1,-M+1\right)
}\left( 1;a_{n_{2}}^{\left( N,M\right) }\lambda _{2}\right)%
\end{array}%
\right\vert  \notag \\
&=&\left( -1\right) ^{n_{1}+n_{2}}\frac{b_{n_{1}}^{\left( N,M\right)
}b_{n_{2}}^{\left( N,M\right) }}{2(N-1)}\times A,
\end{eqnarray}%
where (see Eqs.~(\ref{eq:a}), (\ref{eq:g}), and (\ref{eq:lambda-lambda})) 
\begin{eqnarray}
A &=&\left\vert 
\begin{array}{cc}
\left( -1\right) ^{n_{1}-N+1}\lambda _{1}+\lambda _{n_{1}}^{\left(
N,M\right) } & \left( -1\right) ^{n_{2}-N+1}\lambda _{2}+\lambda
_{n_{2}}^{\left( N,M\right) } \\ 
n_{1}^{2}a_{n_{1}}^{\left( N,M\right) }\left( \left( -1\right)
^{n_{1}-N+1}\lambda _{1}+\lambda _{n_{1}}^{\left( N,M\right) }\right) & 
n_{2}^{2}a_{n_{2}}^{\left( N,M\right) }\left( \left( -1\right)
^{n_{2}-N+1}\lambda _{2}+\lambda _{n_{2}}^{\left( N,M\right) }\right)%
\end{array}%
\right\vert  \notag \\
&=&\frac{\left( \left( -1\right) ^{n_{1}-N+1}\lambda _{1}+\lambda
_{n_{1}}^{\left( N,M\right) }\right) \left( \left( -1\right)
^{n_{2}-N+1}\lambda _{2}+\lambda _{n_{2}}^{\left( N,M\right) }\right) }{4} 
\notag \\
&&\times \left( E_{-n_{1}-1}\left( N,M\right) -E_{-n_{2}-1}\left( N,M\right)
\right) .
\end{eqnarray}%
Consequently 
\begin{eqnarray}
&&W\left( p_{n_{1}}^{\left( -N,-M\right) }\left( z;\lambda _{1}\right)
,p_{n_{2}}^{\left( -N,-M\right) }\left( z;\lambda _{2}\right) \mid +1\right)
\notag \\
&=&\frac{b_{n_{1}}^{\left( N,M\right) }b_{n_{2}}^{\left( N,M\right) }}{8(N-1)%
}\left( \lambda _{1}+\left( -1\right) ^{n_{1}-N+1}\lambda _{n_{1}}^{\left(
N,M\right) }\right) \left( \lambda _{2}+\left( -1\right) ^{n_{2}-N+1}\lambda
_{n_{2}}^{\left( N,M\right) }\right)  \notag \\
&&\times \left( E_{-n_{1}-1}\left( N,M\right) -E_{-n_{2}-1}\left( N,M\right)
\right) .  \label{eq:W1}
\end{eqnarray}

%
%
By comparing Eq.~(\ref{eq:W-1}) and Eq.~(\ref{eq:W1}), the regularity
condition (\ref{eq:reg-2}) for the two-step extension $V^{\left(
-n_1-1,-n_2-1)\right) }\left( x;N,M;\lambda _{1},\lambda _{2}\right) $ can
be written as 
\begin{equation}
\mathrm{sign}\left( \lambda _{1}\lambda _{2}\right) =\mathrm{sign}\left(
\left( -1\right) ^{n_{1}+n_{2}-1}\left( \lambda _{1}+\left( -1\right)
^{n_{1}-N+1}\lambda _{n_{1}}^{\left( N,M\right) }\right) \left( \lambda
_{2}+\left( -1\right) ^{n_{2}-N+1}\lambda _{n_{2}}^{\left( N,M\right)
}\right) \right) .  \label{eq:reg-two-step-bis}
\end{equation}
Since the regularity condition for the first extension $V^{-n_1-1}\left(
x;N,M;\lambda _{1}\right) $ (see \cite{bagchi}) implies that 
\begin{equation}
\mathrm{sign}\left( \lambda _{1}\right) =\mathrm{sign}\left( \left(
-1\right) ^{n_{1}}g_{n_{1}}^{\left( N,M\right) }\left( \lambda _{1}\right)
\right) =\mathrm{sign}\left( \left( -1\right) ^{n_{1}-N-M+1}\left( \lambda
_{1}+\left( -1\right) ^{n_{1}-N+1}\lambda _{n_{1}}^{\left( N,M\right)
}\right) \right),
\end{equation}
Eq.~(\ref{eq:reg-two-step-bis}) simply becomes 
\begin{equation}
\mathrm{sign}\left( \lambda _{2}\right) =\mathrm{sign}\left( \left(
-1\right) ^{n_{2}-N-M}\lambda _{2}+\left( -1\right) ^{M-1}\lambda
_{n_{2}}^{\left( N,M\right) }\right) .  \label{eq:reg-two-step-ter}
\end{equation}

%
%
Consider first the case where $M$, $n_1-N$, and $n_2-N$ are all even. The
regularity of $V^{\left( -n_{1}-1\right) }\left( x;N,M;\lambda _{1}\right) $
necessitates $0<\lambda _{1}<\lambda _{n_{1}}^{\left( N,M\right) }$. Then
condition (\ref{eq:reg-two-step-ter}) becomes 
\begin{equation}
\mathrm{sign}\left( \lambda _{2}\right) =\mathrm{sign}\left( \lambda
_{2}-\lambda _{n_{2}}^{\left( N,M\right) }\right) .
\end{equation}
This condition is always achieved when $\lambda _{2}<0$ and if $\lambda
_{2}>0$ provided 
\begin{equation}
\lambda _{2}>\lambda _{n_{2}}^{\left( N,M\right) }.
\end{equation}
In the same manner, we find the other regularity conditions. Consequently,
we have the regularity conditions mentioned below: 
\begin{equation}
\left\{ 
\begin{array}{c}
(M,n_1-N,n_2-N)\in \left( 2\mathbb{N}\right)^{3}:0<\lambda_1 <\lambda
_{n_1}^{\left( N,M\right) }, \lambda_2<0 \text{ or } \lambda_2>%
\lambda_{n_2}^{(N,M)}; \\[0.2cm] 
(M,n_1-N,n_2-N)\in \left(2\mathbb{N}\right)^2 \times (2\mathbb{N}%
+1):0<\lambda_1 <\lambda _{n_1} ^{\left( N,M\right) },
-\lambda_{n_2}^{(N,M)}<\lambda_2<0; \\[0.2cm] 
(M,n_1-N,n_2-N)\in 2\mathbb{N} \times (2\mathbb{N}+1) \times 2\mathbb{N}%
:\lambda_1<- \lambda_{n_1}^{(N,M)} \text{ or } \lambda_1>0, \lambda_2<0 \\ 
\text{ or } \lambda_2>\lambda_{n_2}^{(N,M)}; \\[0.2cm] 
(M,n_1-N,n_2-N)\in 2\mathbb{N} \times (2\mathbb{N}+1)^2:
\lambda_1<-\lambda_{n_1}^{(N,M)} \text{ or } \lambda_1>0, \\ 
-\lambda_{n_2}^{(N,M)}<\lambda_2<0; \\[0.2cm] 
(M,n_1-N,n_2-N) \in (2\mathbb{N}+1)\times(2\mathbb{N})^2: \lambda_1<0 \text{
or } \lambda_1> \lambda_{n_1}^{(N,M)}, \\ 
0<\lambda_2<\lambda_{n_2}^{(N,M)}; \\[0.2cm] 
(M,n_1-N,n_2-N) \in (2\mathbb{N}+1)\times 2\mathbb{N} \times (2\mathbb{N}%
+1): \lambda_1<0 \text{ or } \lambda_1>\lambda_{n_1}^{(N,M)}, \\ 
\lambda_2<-\lambda_{n_2}^{(N,M)} \text{ or } \lambda_2>0; \\[0.2cm] 
(M,n_1-N,n_2-N) \in (2\mathbb{N}+1)^2 \times 2\mathbb{N}:
-\lambda_{n_1}^{(N,M)}<\lambda_1<0, 0<\lambda_2<\lambda_{n_2}^{(N,M)}; \\%
[0.2cm] 
(M,n_1-N,n_2-N) \in (2\mathbb{N}+1)^3: -\lambda_{n_1}^{(N,M)}<\lambda_1<0,
\lambda_2< -\lambda_{n_2}^{(N,M)} \text{ or } \lambda_2>0.%
\end{array}
\right.  \label{eq:reg-two-step-quater}
\end{equation}

%
%

\subsection{Rationally-extended potential}

If $\lambda _{1}$ and $\lambda _{2}$ satisfy conditions (\ref%
{eq:reg-two-step-quater}), then $\psi _{-n_{1}-1}\left( x;N,M;\lambda
_{1}\right) $ and $\psi _{-n_{2}-1}\left( x;N,M;\lambda _{2}\right) $ can be
used as seed functions to build a two-step state-adding chain of DT, which
generates a regular rational extension of $V\left( x;N-2,M-2\right) $, 
\begin{align}
& V^{-n_{1}-1,-n_{2}-1}(x;N,M;\lambda _{1},\lambda _{2})  \notag \\
& \quad =V(x;N,M)-2\frac{d^{2}}{dx^{2}}\log W\left( \psi
_{-n_{1}-1}(x;N,M;\lambda _{1}),\psi _{-n_{2}-1}(x;N,M;\lambda _{2})\mid
x\right)  \notag \\
& \quad =V(x;N,M)-2\frac{d^{2}}{dx^{2}}\log \left( \psi _{-1}^{2}(x;N,M)%
\frac{dz}{dx}\right)  \notag \\
& \qquad -2\frac{d^{2}}{dx^{2}}\log W\left( p_{n_{1}}^{(-N,-M)}(z;\lambda
_{1}),p_{n_{2}}^{(-N,-M)}(z;\lambda _{2})\mid z\right)  \notag \\
& \quad =V(x;N-2,M-2)+E_{-2}(N,M)  \notag \\
& \qquad -2\frac{d^{2}}{dx^{2}}\log W\left( p_{n_{1}}^{(-N,-M)}(z;\lambda
_{1}),p_{n_{2}}^{(-N,-M)}(z;\lambda _{2})\mid z\right) ,
\end{align}%
where use has been made of (\ref{wronsk prop}) as well as of the relation 
\begin{equation}
\psi _{-1}^{2}(z;N,M)\frac{dz}{dx}\propto \psi _{-1}(z;N,M)\psi
_{-1}(z;N-1,M-1),
\end{equation}%
the shape invariance property of the TDPT potential, 
\begin{equation}
V(x;N,M)-2\frac{d^{2}}{dx^{2}}\log \psi
_{-1}(z;N,M)=V(x;N-1,M-1)+E_{-1}(N,M),  \label{eq:shape}
\end{equation}%
and the relation 
\begin{equation}
E_{-2}(N,M)=E_{-1}(N,M)+E_{-1}(N-1,M-1).
\end{equation}

%
%
It only remains to make the change of variable from $x$ to $z$ to get the
final result 
\begin{align}
& V^{(-n_1-1,-n_2-1)}(x;N,M;\lambda_1,\lambda_2)  \notag \\
&\quad = V(x;N-2,M-2) + E_{-2}(N,M)  \notag \\
&\qquad -8(1-z^2) \frac{d^2}{dz^2} \log
W\left(p_{n_1}^{(-N,-M)}(z;\lambda_1), p_{n_2}^{(-N,-M)} (z;\lambda_2) \mid
z\right)  \notag \\
&\qquad + 8z\frac{d}{dz} \log W\left(p_{n_1}^{(-N,-M)}(z;\lambda_1),
p_{n_2}^{(-N,-M)} (z;\lambda_2) \mid z\right).  \label{eq:V-ext}
\end{align}
For $V^{(-n_1-1,-n_2-1)}(x;N,M;\lambda_1,\lambda_2)$ to be a confining
potential on $]0,\pi/2[$, we need to impose this property to $V(x;N-2,M-2)$.
This is achieved for $N,M\ge 3$. For such a choice and provided $n_1$ and $%
n_2$ satisfy Eq.~(\ref{eq:n_1-n_2}), $V^{(-n_1-1,-n_2-1)}(x;N,M;\lambda_1,%
\lambda_2)$ is strongly repulsive in both 0 and $\pi/2$, so that in each
extremity only one basis solution is quadratically integrable and the
corresponding Hamiltonian is essentially self-adjoint \cite{frank}.

%
%
The corresponding eigenstates are given by 
\begin{align}
  & \psi_k^{(-n_1-1,-n_2-1)}(x;N,M;\lambda_1,\lambda_2)  \notag \\
  &\quad = \frac{W(\psi_{-n_1-1}(x;N,M;\lambda_1,\lambda_2),
\psi_{-n_2-1}(x;N,M;\lambda_1,\lambda_2), \psi_k(x;N,M) \mid x)}{%
W(\psi_{-n_1-1}(x;N,M;\lambda_1,\lambda_2), \psi_{-n_2-1}(x;N,M;
\lambda_1,\lambda_2) \mid x)},  \label{eq:psi-1}
\end{align}
for $k=0, 1, 2, \ldots$, as well as 
\begin{align}
& \psi_{-n_1-1}^{(-n_1-1,-n_2-1)}(x;N,M;\lambda_1,\lambda_2)  \notag \\
&\quad = \frac{\psi_{-n_2-1}(x;N,M;\lambda_1,\lambda_2)}{W(%
\psi_{-n_1-1}(x;N,M;\lambda_1,\lambda_2),
\psi_{-n_2-1}(x;N,M;\lambda_1,\lambda_2) \mid x)},  \label{eq:psi-2}
\end{align}
and 
\begin{align}
& \psi_{-n_2-1}^{(-n_1-1,-n_2-1)}(x;N,M;\lambda_1,\lambda_2)  \notag \\
&\quad = \frac{\psi_{-n_1-1}(x;N,M;\lambda_1,\lambda_2)}{W(%
\psi_{-n_1-1}(x;N,M;\lambda_1,\lambda_2),
\psi_{-n_2-1}(x;N,M;\lambda_1,\lambda_2) \mid x)}.  \label{eq:psi-3}
\end{align}
The corresponding energies are $E_k(N,M)$ ($k=0,1,2,\ldots$), $%
E_{-n_1-1}(N,M)$, and $E_{-n_2-1}(N,M)$, respectively.
\par
%
%
On using (\ref{wronsk prop}) again, as well as derivative properties of
Jacobi and para-Jacobi polynomials, Eqs.~(\ref{eq:psi-1}), (\ref{eq:psi-2}),
and (\ref{eq:psi-3}) can be rewritten as 
\begin{equation}
  \psi _{k}^{(-n_{1}-1,-n_{2}-1)}(x;N,M;\lambda _{1},\lambda _{2})\propto \psi
  _{0}(x;N-2,M-2)\frac{Q_{k}^{(n_{1},n_{2})}(z;N,M;\lambda _{1},\lambda _{2})}{%
  T_{n_{1}n_{2}}^{(-N,-M)}(z;\lambda _{1},\lambda _{2})},
\end{equation}
where 
\begin{align}
  & Q_{k}^{(n_{1},n_{2})}(z;N,M;\lambda _{1},\lambda
    _{2})=4(k+1)(k+2)T_{n_{1}n_{2}}^{(-N,-M)}(z;\lambda _{1},\lambda _{2})P_{k+2}^{(N-2,M-2)}(z) 
    \notag \\
  & \quad -2(k+1)U_{n_{1}n_{2}}^{(-N,-M)}(z;\lambda _{1},\lambda
    _{2})(1-z^{2})P_{k+1}^{(N-1,M-1)}(z)  \notag \\
  & \quad +V_{n_{1}n_{2}}^{(-N,-M)}(z;\lambda _{1},\lambda
    _{2})(1-z^{2})^{2}P_{k}^{(N,M)}(z),\qquad k=0,1,2,\ldots , \\
  & Q_{-n_{1}-1}^{(n_{1},n_{2})}(z;N,M;\lambda _{1},\lambda
    _{2})=p_{n_{2}}^{(-N,-M)}(z;\lambda _{2}), \\
  & Q_{-n_{2}-1}^{(n_{1},n_{2})}(z;N,M;\lambda _{1},\lambda
    _{2})=p_{n_{1}}^{(-N,-M)}(z;\lambda _{1}).
\end{align}
In these equations, we have defined 
\begin{align}
& T_{n_{1}n_{2}}^{(-N,-M)}(z;\lambda _{1},\lambda _{2})  \notag \\
& \quad =n_{2}p_{n_{1}}^{(-N,-M)}(z;\lambda
_{1})p_{n_{2}-1}^{(-N+1,-M+1)}\left( z;a_{n_{2}}^{(N,M)}\lambda _{2}\right) 
\notag \\
& \qquad -n_{1}p_{n_{1}-1}^{(-N+1,-M+1)}\left( z;a_{n_{1}}^{(N,M)}\lambda
_{1}\right) p_{n_{2}}^{(-N,-M)}(z;\lambda _{2}), \\
& U_{n_{1}n_{2}}^{(-N,-M)}(z;\lambda _{1},\lambda _{2})  \notag \\
& \quad =n_{2}(n_{2}-1)p_{n_{1}}^{(-N,-M)}(z;\lambda
_{1})p_{n_{2}-2}^{(-N+2,-M+2)}\left(
z;a_{n_{2}-1}^{(N-1,M-1)}a_{n_{2}}^{(N,M)}\lambda _{2}\right)  \notag \\
& \qquad -n_{1}(n_{1}-1)p_{n_{1}-2}^{(-N+2,-M+2)}\left(
z;a_{n_{1}-1}^{(N-1,M-1)}a_{n_{1}}^{(N,M)}\lambda _{1}\right)
p_{n_{2}}^{(-N,-M)}(z;\lambda _{2}), \\
& V_{n_{1}n_{2}}^{(-N,-M)}(z;\lambda _{1},\lambda _{2})  \notag \\
& \quad =n_{1}n_{2}(n_{2}-1)p_{n_{1}-1}^{(-N+1,-M+1)}\left(
z;a_{n_{1}}^{(N,M)}\lambda _{1}\right) p_{n_{2}-2}^{(-N+2,-M+2)}\left(
z;a_{n_{2}-1}^{(N-1,M-1)}a_{n_{2}}^{(N,M)}\lambda _{2}\right)  \notag \\
& \qquad -n_{1}(n_{1}-1)n_{2}p_{n_{1}-2}^{(-N+2,-M+2)}\left(
z;a_{n_{1}-1}^{(N-1,M-1)}a_{n_{1}}^{(N,M)}\lambda _{1}\right)  \notag \\
& \qquad \times p_{n_{2}-1}^{(-N+1,-M+1)}\left( z;a_{n_{2}}^{(N,M)}\lambda
_{2}\right).
\end{align}
\par
%
%
Due to the orthogonality properties of the $\psi_k^{(-n_1-1,-n_2-1)}(x;N,M;%
\lambda_1,\lambda_2)$, we deduce that the $Q_k^{(n_1,n_2)}(z;N,M;\lambda_1,%
\lambda_2)$, $k=-n_1-1, -n_2-1, 0, 1, 2, \ldots$, are orthogonal polynomials
on $]-1,1[$ with respect to the measure 
\begin{equation}
\mu_{n_1n_2}^{(-N,-M)}(z;\lambda_1,\lambda_2) = \frac{(1-z)^{N-2} (1+z)^{M-2}%
}{\left(T_{n_1n_2}^{(-N,-M)}(z;\lambda_1,\lambda_2)\right)^2}.
\end{equation}
%
%

\subsection{Explicit examples}

Consider for instance the case where $N=M=3$. The seed function indices $%
\left( n_{1},n_{2}\right) $ of the two-step chain can be chosen in the set ($%
3\leq n_{2}<n_{1}<6$), so that $(n_1,n_2) \in\left\{ \left( 4,3\right)
,(5,4),(5,3)\right\} $. As for the possible values of the regularity
parameters $\lambda _{n_{i}}^{\left( N,M\right) },\ i=1,2$, we get from Eq.~(%
\ref{eq:lambda}) 
\begin{equation}
\lambda _{3}^{\left( 3,3\right) }=3,\ \lambda _{4}^{\left( 3,3\right) }=2,\
\lambda _{5}^{\left( 3,3\right) }=1/6.\ 
\end{equation}
From Eq.~(\ref{eq:para-Jacobi}), the corresponding para-Jacobi polynomials
can be written as 
\begin{equation}
\left\{ 
\begin{array}{c}
p_{3}^{\left( -3,-3\right) }\left( z;\lambda \right) =z^{3}+(3-2\lambda
)z^{2}+3z+1-\frac{2}{3}\lambda, \\[0.1cm] 
p_{4}^{\left( -3,-3\right) }\left( z;\lambda \right) =z^{4}-6z^{2}-8(\lambda
+1)z-3, \\[0.1cm] 
p_{5}^{\left( -3,-3\right) }\left( z;\lambda \right) =z^{5}-\frac{10}{3}
z^{3}+5z+8(\frac{1}{3}-4\lambda ).%
\end{array}
\right.
\end{equation}

%
%
Then, from Eq.~(\ref{eq:V-ext}), 
\begin{eqnarray}
V^{\left( -n_{1}-1,-n_{2}-1\right) }\left( x;3,3;\lambda _{1},\lambda
_{2}\right) &=& V(x;1,1)+E_{-2}(3,3)  \notag \\
&&{}+\Delta V^{\left( -n_{1}-1,-n_{2}-1\right) }\left( x;3,3;\lambda
_{1},\lambda _{2}\right) ,
\end{eqnarray}
with 
\begin{equation}
V(x;1,1)=\frac{3}{1-z^{2}}-9,\qquad E_{-2}(3,3)=-40,
\end{equation}
and 
\begin{eqnarray}
&& \Delta V^{\left( -n_{1}-1,-n_{2}-1\right) }\left( x;3,3;\lambda
_{1},\lambda _{2}\right) = 8z\frac{d}{dz}\log W\left( p_{n_{1}}^{\left(
-3,-3\right) }\left( z;\lambda _{1}\right) ,p_{n_{2}}^{\left( -3,-3\right)
}\left( z;\lambda _{2}\right) \mid z\right)  \notag \\
&&\qquad {}-8(1-z^{2})\frac{d^{2}}{dz^{2}}\log W\left( p_{n_{1}}^{\left(
-3,-3\right) }\left( z;\lambda _{1}\right) ,p_{n_{2}}^{\left( -3,-3\right)
}\left( z;\lambda _{2}\right) \mid z\right) .
\end{eqnarray}

%
%
If $n_1=4$ and $n_2=3$, for instance, 
\begin{eqnarray}
&& \Omega \left( z;\lambda _{1},\lambda _{2}\right)  \notag \\
&& \qquad = W\left(p_{n_{1}}^{\left( -3,-3\right) }\left( z;\lambda
_{1}\right) ,p_{n_{2}}^{\left( -3,-3\right) }\left( z;\lambda _{2}\right)
\mid z\right)  \notag \\
&& \qquad = \left\vert 
\begin{array}{cc}
z^{4}-6z^{2}-8(\lambda _{1}+1)z-3 & z^{3}+(3-2\lambda _{2})z^{2}+3z+1-\frac{%
2 }{3}\lambda _{2} \\ 
4z^{3}-12z-8(\lambda _{1}+1) & 3z^{2}+2(3-2\lambda _{2})z+3%
\end{array}%
\right\vert  \notag \\
&& \qquad = -z^{6}+\left( 4\lambda _{2}-6\right) z^{5}-15z^{4}+\left( \frac{8%
}{3} \lambda _{2}-16\lambda _{1}-20\right) z^{3}  \notag \\
&& \qquad \quad +(16\lambda _{2}\lambda _{1}-24\lambda _{1}+16\lambda
_{2}-15)z^{2}+\left( 4\lambda _{2}-6\right) z+8\lambda _{1}-\frac{16}{3}%
\lambda _{2}  \notag \\
&& \qquad \quad -\frac{16}{3}\lambda _{1}\lambda _{2}-1
\end{eqnarray}
and 
\begin{eqnarray}
&&V^{\left( -5,-4\right) }\left( z;3,3;\lambda _{1},\lambda _{2}\right) 
\notag \\
&&\qquad =\frac{3}{1-z^{2}}-17+32(1-z^{2})\left( \frac{A\left( z;\lambda
_{1},\lambda _{2}\right) }{\Omega \left( z;\lambda _{1},\lambda _{2}\right) }%
\right)^{2} -16\frac{B\left( z;\lambda _{1},\lambda _{2}\right) }{\Omega
\left( z;\lambda _{1},\lambda _{2}\right) },  \label{eq:example-V}
\end{eqnarray}
with 
\begin{equation}
\left\{ 
\begin{array}{c}
A\left( z;\lambda _{1},\lambda _{2}\right) =3z^{5}+\left( 15-10\lambda
_{2}\right) z^{4}+30z^{3}+\left( 24\lambda _{1}-4\lambda _{2}+30\right) z^{2}
\\ 
+(24\lambda _{1}-16\lambda _{2}-16\lambda _{2}\lambda _{1}+15)z-\left(
2\lambda _{2}-3\right), \\[0.1cm] 
B\left( z;\lambda _{1},\lambda _{2}\right) =18z^{6}-\left( 50\lambda
_{2}-75\right) z^{5}+105z^{4}+\left( 28\lambda _{2}+72\lambda
_{1}+30\right)z^{3} \\ 
+(48\lambda _{1}-32\lambda _{2}-32\lambda _{2}\lambda _{1}-60)z^{2} +\left(
6\lambda _{2}-48\lambda _{1}-57\right) z \\ 
+(16\lambda _{2}-24\lambda _{1}+16\lambda _{1}\lambda _{2}-15).%
\end{array}
\right.
\end{equation}
\par
%
%
As for the regularity conditions, they are given by 
\begin{equation}
\left\{ 
\begin{array}{c}
-2<\lambda _{1}<0, \\ 
0<\lambda _{2}<3.%
\end{array}
\right.
\end{equation}
\par
%
%
In Fig.~1, potential (\ref{eq:example-V}) is plotted in terms of $z$ for $\lambda_1=-1$ and $\lambda_2=1$. It is also displayed in terms of $z$ and $t=\lambda_1$ for $\lambda_2=1$ or in terms of $z$ and $\mu=\lambda_2$ for $\lambda_1 =-1$ in Figs.~2 and 3, respectively.

\begin{center}
\includegraphics[scale=0.8]{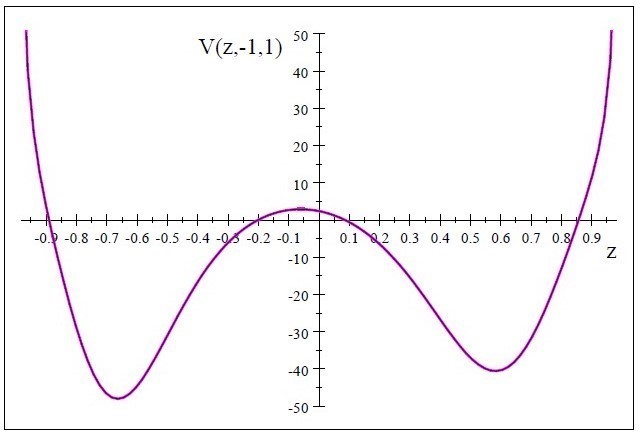}
\captionof{figure}{Potential $V^{(-5,-4)}(z;3,3;-1,1)$ in terms of $z$ for $-1<z<1$.}
\label{fig1}
\end{center}

\begin{center}
\includegraphics[scale=0.4]{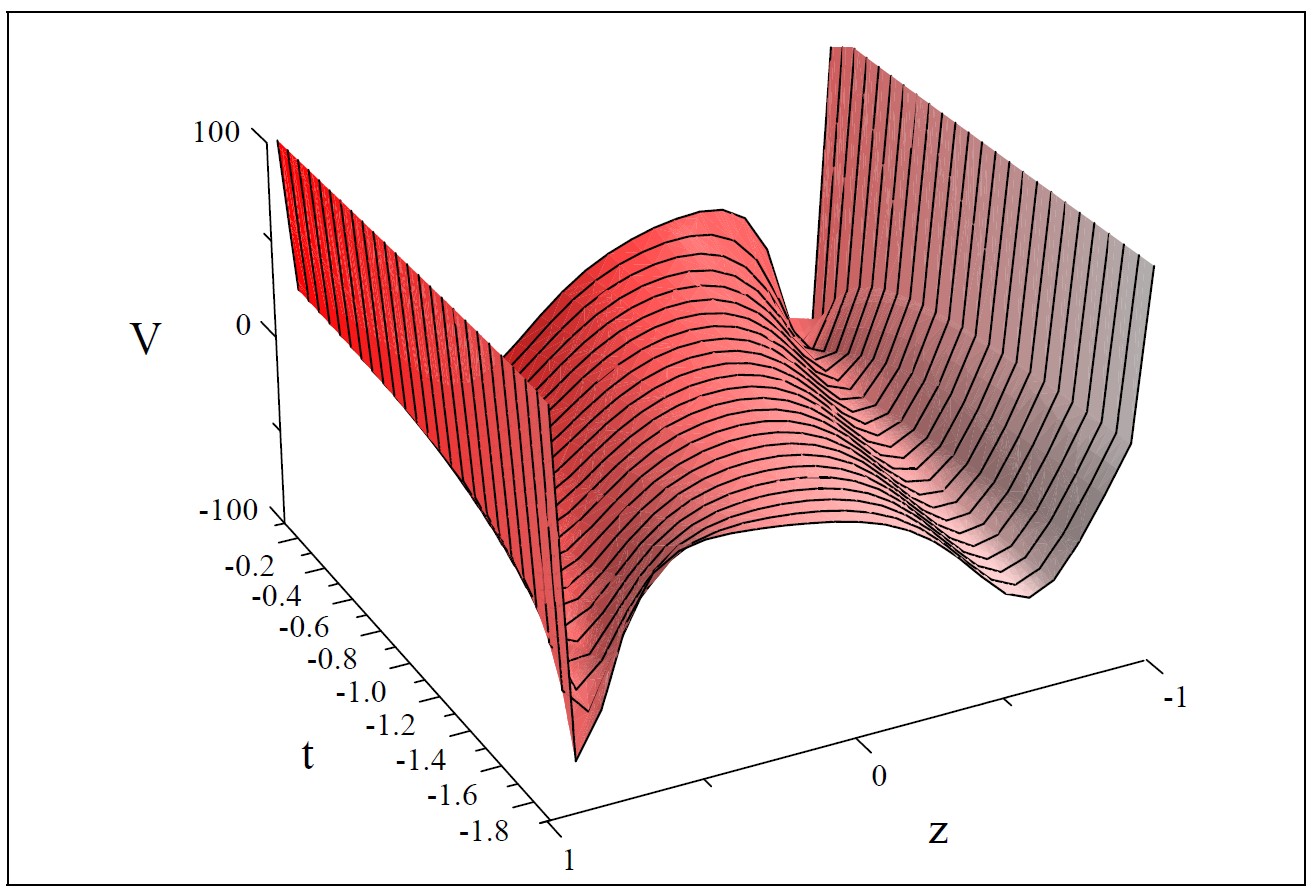}
\captionof{figure}{Potential $V^{(-5,-4)}(z;3,3;t,1)$ in terms of $z$ and $t$ for $-1<z<1$ and $-2<t<0$.}
\label{fig2}
\end{center}

\begin{center}
\includegraphics[scale=0.4]{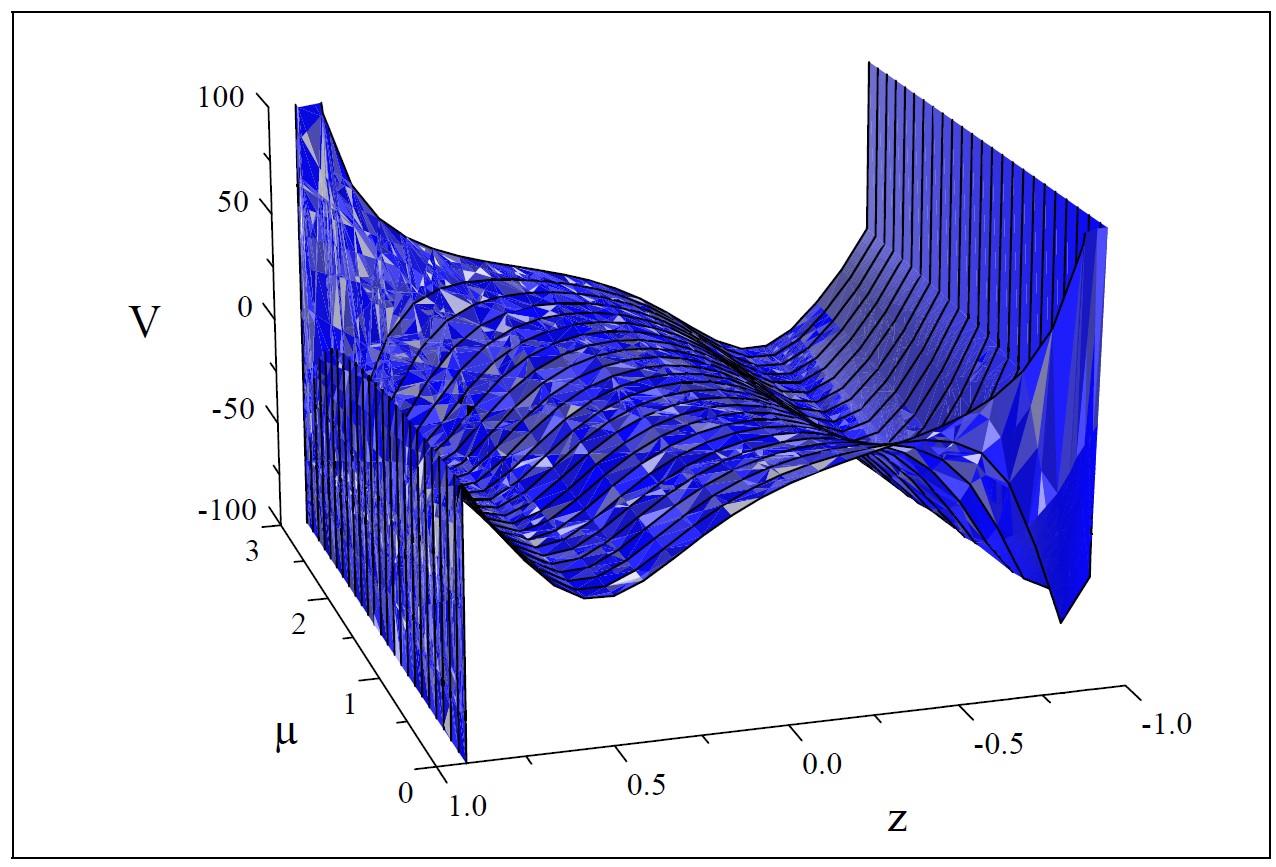}
\captionof{figure}{Potential $V^{(-5,-4)}(z;3,3;-1,\mu)$ in terms of $z$ and $\mu$ for $-1<z<1$ and $0<\mu<3$.}
\label{fig3}
\end{center}
\par
%
%
\section{\boldmath $m$-STEP REGULAR EXTENSIONS OF THE TDPT POTENTIAL}

\subsection{Regularity conditions}

\setcounter{equation}{0}

Consider now an $m$-step chain $N_m = (-n_1-1, -n_2-1, \ldots, -n_m-1)$ of
DT based on PJ seed functions $\psi_{-n_{1}-1 }( x;N,M;\lambda _{1}) $, $%
\psi_{-n_{2}-1 }( x;N,M;\lambda _{2}) $, \ldots, $\psi_{-n_m-1}(x;N,M;%
\lambda_m)$, 
\begin{align}
&V\left( x;N,M\right) \overset{A(\psi _{-n_1-1})}{ \longrightarrow }%
V^{\left( -n_{1}-1\right) }\left( x;N,M;\lambda _{1}\right) \overset{%
A\left(\psi^{(-n_1-1)}_{-n_2-1}\right) }{\longrightarrow }\cdots  \notag \\
& \overset{A\left(\psi^{(N_m-2)}_{-n_{m-1}-1}\right)}{\longrightarrow }%
V^{(N_{m-1})}(x;N,M;\lambda_1, \ldots, \lambda_{m-1}) \overset{%
A\left(\psi^{(N_{m-1})} _{-n_m-1}\right)}{\longrightarrow}
V^{(N_m)}(x;N,M;\lambda_1,\ldots, \lambda_m),
\end{align}
where 
\begin{align}
&\psi^{(N_{m-1})}_{-n_m-1}(x;N,M;\lambda_1,\ldots, \lambda_m)  \notag \\
&\quad = \frac{W(\psi_{-n_1-1}(x;N,M;\lambda_1),
\psi_{-n_2-1}(x;N,M;\lambda_2), \ldots, \psi_{-n_m-1}(x;N,M;\lambda_m) \mid x)%
} {W(\psi_{-n_1-1}(x;N,M;\lambda_1), \psi_{-n_2-1}(x;N,M;\lambda_2), \ldots,
\psi_{-n_{m-1}-1}(x;N,M;\lambda_{m-1}) \mid x)}.
\end{align}

%
%
On assuming that the extension obtained at the $(m-1)$th step is regular,
which imposes that $\max \left( N,M\right) \leq n_{m-1}<\cdots
<n_{2}<n_{1}<N+M$ and definite conditions on $\lambda _{1},\lambda
_{2},\ldots ,\lambda _{m-1}$, we want to impose that the same is true at the 
$m$th step. For this to occur, $\psi^{(N_{m-1})}_{-n_{m}-1}$ must be in the
disconjugacy sector of $V^{(N_{m-1})}$, i.e., $\max \left( N,M\right) \leq
n_{m}<n_{m-1}<\cdots <n_{1}<N+M$, and, in addition, it must not change sign
on $]0,\pi /2[$, which leads to the condition 
\begin{align}
  & \mathrm{sign}(W(\psi _{-n_{1}-1},\psi _{-n_{2}-1},\ldots ,\psi
    _{-n_{m}-1}\mid 0))  \notag \\
  & \quad =\mathrm{sign}(W(\psi _{-n_{1}-1},\psi _{-n_{2}-1},\ldots ,\psi
    _{-n_{m}-1}\mid \pi /2)).  \label{eq:reg-m}
\end{align}
Since standard properties of Wronskians (\ref{wronsk prop}) lead to 
\begin{align}
  & W(\psi _{-n_{1}-1},\psi _{-n_{2}-1},\ldots ,\psi _{-n_{m}-1}\mid x)\propto
    (1-z)^{\frac{m}{2}\left( -N+\frac{m}{2}\right) }(1+z)^{\frac{m}{2}\left( -M+%
    \frac{m}{2}\right) }  \notag \\
  & \qquad \times W\left( p_{n_{1}}^{(-N,-M)}(z;\lambda
    _{1}),p_{n_{2}}^{(-N,-M)}(z;\lambda _{2}),\ldots
    ,p_{n_{m}}^{(-N,-M)}(z;\lambda _{m})\mid z\right) ,
\end{align}
condition (\ref{eq:reg-m}) can be rewritten as
\begin{eqnarray}
  && {\rm sign}\left( W\left( p_{n_{1}}^{\left( -N,-M\right) }\left( z;\lambda
    _{1}\right) ,\ldots,p_{n_{m}}^{\left( -N,-M\right) }\left( z;\lambda
    _{m}\right) \mid 1\right) \right) \nonumber \\
  && \quad = {\rm sign}\left( W\left( p_{n_{1}}^{\left(
    -N,-M\right) }\left( z;\lambda _{1}\right) ,\ldots,p_{n_{m}}^{\left(
    -N,-M\right) }\left( z;\lambda _{m}\right) \mid -1\right) \right) .
\end{eqnarray}
\par
%
%
The derivation rule for the para-Jacobi polynomials (\ref{eq:para-der})
leads to 
\begin{equation}
  \frac{d^{k}}{dz^{k}}p_{n_{j}}^{\left( -N,-M\right) }\left( z;\lambda \right)
  =\left( n_{j}\right) _{\underline{k}}p_{n_{j}-k}^{\left( -N+k,-M+k\right)
  }\left( z;A_{j}^{(k)}\lambda \right) ,
\end{equation}
where (cf Eq.~(\ref{eq:a}))
\begin{equation}
  A_{j}^{(k)}=\prod_{i=0}^{k-1}a_{n_{j}-i}^{\left( N-i,M-i\right) }=\frac{%
  \left( N+M-n_{i}-1\right) _{\underline{k}}}{\left( n\right) _{\underline{k}}},  \label{eq:A}
\end{equation}
$(x)_{\underline{p}}$ being the falling factorial
\begin{equation}
(x)_{\underline{l}}=x(x-1)...(x-l+1).  \label{fall fact}
\end{equation}
\par
%
%
This gives
\begin{equation}
  W\left( p_{n_{1}}^{(-N,-M)}(z;\lambda _{1}),p_{n_{2}}^{(-N,-M)}(z;\lambda
  _{2}),\ldots ,p_{n_{m}}^{(-N,-M)}(z;\lambda _{m})\mid z\right) =\det \left[ 
  \mathbf{R}\left( z;\lambda _{1},...,\lambda _{m}\right) \right] ,
  \label{mat R}
\end{equation}
where the elements of the matrix are given by
\begin{equation}
  \mathbf{R}_{i,j}\left( z;\lambda _{1},...,\lambda _{m}\right) =\left(
  n_{j}\right) _{\underline{i-1}}p_{n_{j}-i+1}^{\left( -N+i-1,-M+i-1\right)
  }\left( z;A_{j}^{(i-1)}\lambda _{j}\right) .  \label{mat R 2}
\end{equation}
\par
%
%
Condition (\ref{eq:reg-m}) can then be rewritten as 
\begin{equation}
  \mathrm{sign}\left( \det \left[\mathbf{R}\left( 1;\lambda _{1},...,\lambda _{m}\right) \right] \right) 
  =\mathrm{sign}\left( \det \left[\mathbf{R}\left( -1;\lambda _{1},...,\lambda _{m}\right) \right] \right) .
  \label{eq:reg-m-step}
\end{equation}
\par
%
%
Consider first the right-hand side of Eq.~(\ref{eq:reg-m-step}). By using
Eqs.~(\ref{eq:para-1}), (\ref{eq:b}), and since, from (\ref{eq:b-b}), 
\begin{equation}
b_{n-k}^{\left( N-k,M-k\right) }=\frac{\left( n\right) _{\underline{k}}}{%
2^{k}(M-1)_{\underline{k}}}b_{n}^{\left( N,M\right) },
\end{equation}
we obtain for $\det \left[\mathbf{R}_{i,j}\left( -1;\lambda _{1},...,\lambda
_{m}\right) \right] $, 
\begin{eqnarray}
&&\left\vert 
\begin{array}{ccc}
\lambda _{1}\left( -1\right) ^{n_{1}}b_{n_{1}}^{\left( N,M\right) } & ... & 
\lambda _{m}\left( -1\right) ^{n_{m}}b_{n_{m}}^{\left( N,M\right) } \\ 
... &  & ... \\ 
\lambda _{1}\left( -1\right) ^{n_{1}-k}A_{1}^{(k)}\frac{\left( \left(
n_{1}\right) _{\underline{k}}\right) ^{2}}{2^{k}(M-1)_{\underline{k}}}%
b_{n_{1}}^{\left( N,M\right) } & ... & \lambda _{m}\left( -1\right)
^{n_{m}-k}A_{m}^{(k)}\frac{\left( \left( n_{m}\right) _{\underline{k}%
}\right) ^{2}}{2^{k}(M-1)_{\underline{k}}}b_{n_{m}}^{\left( N,M\right) } \\ 
... &  & ... \\ 
\lambda _{1}\left( -1\right) ^{n_{1}-m+1}A_{1}^{(m-1)}\frac{\left( \left(
n_{1}\right) _{\underline{m-1}}\right) ^{2}}{2^{m-1}(M-1)_{\underline{m-1}}}%
b_{n_{1}}^{\left( N,M\right) } & ... & \lambda _{m}\left( -1\right)
^{n_{m}-m+1}A_{m}^{(m-1)}\frac{\left( \left( n_{m}\right) _{\underline{m-1}%
}\right) ^{2}}{2^{m-1}(M-1)_{\underline{m-1}}}b_{n_{m}}^{\left( N,M\right) }%
\end{array}%
\right\vert  \notag \\
&=&\lambda _{1}...\lambda _{m}\frac{\left( -1\right)
^{n_{1}+...+n_{m}+1+...+\left( m-1\right) }b_{n_{1}}^{\left( N,M\right)
}...b_{n_{m}}^{\left( N,M\right) }}{2^{1+...+\left( m-1\right) }\left(
M-1\right) ^{m-1}\left( M-2\right) ^{m-2}...(M-m+1)}D(n_{1},...,n_{m}),
\end{eqnarray}
where 
\begin{eqnarray*}
&&D(n_{1},...,n_{m}) \\
&&\quad=\left\vert 
\begin{array}{ccc}
1 & ... & 1 \\ 
... &  & ... \\ 
A_{1}^{(k)}\left( \left( n_{1}\right) _{\underline{k}}\right) ^{2} & ... & 
A_{m}^{(k)}(\left( n_{m}\right) _{\underline{k}})^{2} \\ 
... &  & ... \\ 
A_{1}^{(m-1)}(\left( n_{1}\right) _{\underline{m-1}})^{2} & ... & 
A_{m}^{(m-1)}(\left( n_{m}\right) _{\underline{m-1}})^{2}%
\end{array}%
\right\vert .
\end{eqnarray*}
Then 
\begin{align}
  &\det \left[\mathbf{R}_{ij}\left( -1;\lambda _{1},...,\lambda _{m}\right) \right] \nonumber \\
  &\quad=\lambda _{1}...\lambda _{m}\frac{\left( -1\right) ^{n_{1}+...+n_{m}+m\left(
     m-1\right) /2}b_{n_{1}}^{\left( N,M\right) }...b_{n_{m}}^{\left( N,M\right) }%
     }{2^{m\left( m-1\right) /2}\left( M-1\right) ^{m-1}\left( M-2\right)
     ^{m-2}...(M-m+1)}D(n_{1},...,n_{m}),  \label{eq:W-1-m}
\end{align}
with 
\begin{eqnarray}
  &&D(n_{1},...,n_{m})  \notag \\
  &&\quad =\left\vert 
    \begin{array}{ccc}
    1 & ... & 1 \\ 
    ... &  & ... \\ 
    \frac{n_{1}!M+N-n_{1}-1)!}{\left( n_{1}-k\right) !(M+N-n_{1}-k-1)!} & ... & 
    \frac{n_{m}!M+N-n_{m}-1)!}{\left( n_{m}-k\right) !(M+N-n_{m}-k-1)!} \\ 
    ... &  & ... \\ 
    \frac{n_{1}!M+N-n_{1}-1)!}{\left( n_{1}-m+1\right) !(M+N-n_{1}-m)!} & ... & 
    \frac{n_{m}!M+N-n_{m}-1)!}{\left( n_{m}-m+1\right) !(M+N-n_{m}-m)!}%
    \end{array}%
  \right\vert .  \label{eq:D}
\end{eqnarray}
\par
%
%
Let us now consider $\det \left[\mathbf{R}_{ij}\left( 1;\lambda _{1},...,\lambda
_{m}\right) \right] $. From Eqs.~(\ref{eq:g}) and (\ref{eq:para1}), we
have
\begin{equation}
  p_{n-k}^{\left( -N+k,-M+k\right) }\left( 1;\lambda \right) =b_{n-k}^{\left(
  N-k,M-k\right) }\left( \left( -1\right) ^{N+M-1}\lambda +\left( -1\right)
  ^{n-M}\lambda _{n-k}^{\left( N-k,M-k\right) }\right)
\end{equation}
and from Eqs.~(\ref{eq:para-der}) and (\ref{eq:A}), we obtain 
  \begin{equation}
  \frac{d^{k}}{dz^{k}}p_{n_{i}}^{\left( -N,-M\right) }\left( 1;\lambda \right)
  =\left( n_{i}\right) _{\underline{k}}b_{n_{i}-k}^{\left( N-k,M-k\right)
  }\left( -1\right) ^{n_{i}-M}\left( \left( -1\right)
  ^{n_{i}-N+1}A_{i}^{(k)}\lambda +\lambda _{n_{i}-k}^{\left( N-k,M-k\right)
  }\right) .
\end{equation}
With (cf Eq.~(\ref{eq:lambda-lambda})) 
\begin{equation}
  \lambda _{n_{i}-k}^{\left( N-k,M-k\right) }=A_{i}^{(k)}\lambda
  _{n_{i}}^{\left( N,M\right) },
\end{equation}
we therefore arrive at 
\begin{equation}
\frac{d^{k}}{dz^{k}}p_{n_i}^{\left( -N,-M\right) }\left( 1;\lambda \right)
=\left( n_{i}\right) _{\underline{k}}b_{n_{i}-k}^{\left( N-k,M-k\right)
}A_{i}^{(k)}\left( -1\right) ^{n_{i}-M}\left( \left( -1\right)
^{n_{i}-N+1}\lambda +\lambda _{n_{i}}^{\left( N,M\right) }\right) .
\end{equation}
\par
%
%
It results that 
\begin{align}
  &\det \left[ \mathbf{R}\left( 1;\lambda _{1},...,\lambda _{m}\right) \right]. \nonumber \\
  &\quad=\det \left( \left( n_{i}\right) _{\underline{k-1}}b_{n_{j}-k+1}^{\left(
     N-k+1,M-k+1\right) }A_{i}^{(k-1)}\times \left( -1\right) ^{M-n_{j}}\left(
     \left( -1\right) ^{n_{j}-N+1}\lambda _{j}+\lambda _{n_{j}}^{\left(
     N,M\right) }\right) \right) ,
\end{align}
This simplifies into 
\begin{eqnarray}
  &&\det \left[ \mathbf{R}\left( 1;\lambda _{1},...,\lambda _{m}\right) \right]
     \notag \\
  &&\quad =\left( -1\right) ^{n_{1}+...+n_{m}-mM}\left( \left( -1\right)
     ^{n_{1}-N+1}\lambda _{1}+\lambda _{n_{1}}^{\left( N,M\right) }\right)
     ...\left( \left( -1\right) ^{n_{m}-N+1}\lambda _{m}+\lambda _{n_{m}}^{\left(
     N,M\right) }\right)  \notag \\
  &&\qquad \times \frac{b_{n_{1}}^{\left( N,M\right) }...b_{n_{m}}^{\left(
     N,M\right) }}{2^{m\left( m-1\right) /2}\left( M-1\right) ^{m-1}\left(
     M-2\right) ^{m-2}...(M-m+1)}D(n_{1},...,n_{m}),  \label{eq:W1-bis}
\end{eqnarray}
with $D(n_{1},...,n_{m})$ given by Eq.~(\ref{eq:D}).
\par
%
%
Comparing Eq.~(\ref{eq:W1-bis}) with Eq.~(\ref{eq:W-1-m}), condition (\ref%
{eq:reg-m-step}) becomes 
\begin{equation}
\mathrm{sign}\left( \lambda _{1}...\lambda _{m}\right) =
(-1)^{m\left(m-1\right) /2} \prod_{j=1}^m \mathrm{sign}%
\left((-1)^{n_j-N-M+1} \lambda_j + (-1)^M \lambda_{n_j}^{(N,M)}\right).
\end{equation}

%
%
As we have assumed that the ($m-1$)-step chain $N_{m-1}=\left(
-n_1-1,\ldots,-n_{{m-1}}-1\right) $ is regular at each level, the condition
above then implies that 
\begin{equation}
\mathrm{sign}\left( \lambda _{1}...\lambda _{m-1}\right) =
(-1)^{(m-1)\left(m-2\right) /2} \prod_{j=1}^{m-1} \mathrm{sign}%
\left((-1)^{n_j-N-M+1} \lambda_j + (-1)^M \lambda_{n_j}^{(N,M)}\right).
\end{equation}
so that the regularity condition for the $m$-step chain reduces to 
\begin{equation}
\mathrm{sign}\left( \lambda _{m}\right) = (-1)^{m-1}\mathrm{sign}\left(
\left( -1\right) ^{n_{m}-N-M+1}\lambda _{m}+\left( -1\right) ^{M}\lambda
_{n_{m}}^{\left( N,M\right) }\right) .
\end{equation}

%
%
Consider first the case where $m$ is even, which leads to 
\begin{equation}
\mathrm{sign}\left( \lambda _{m}\right) =\mathrm{sign}\left( (-1)^{M}\left(
\left( -1\right) ^{n_{m}-N}\lambda _{m}-\lambda _{n_{m}}^{\left( N,M\right)
}\right) \right) .
\end{equation}
We then obtain the following possibilities: 
\begin{equation}
\left\{ 
\begin{array}{c}
(M,n_m-N)\in \left( 2\mathbb{N}\right) ^{2}: \lambda_m>\lambda_{n_m}^{(N,M)} 
\text{ or } \lambda_m<0; \\[0.2cm] 
(M,n_m-N)\in 2\mathbb{N}\times \left( 2\mathbb{N}+1\right) :
-\lambda_m^{(N,M)}<\lambda_m<0; \\[0.2cm] 
(M,n_m-N)\in \left( 2\mathbb{N}+1\right) \times 2\mathbb{N}:
0<\lambda_m<\lambda_{n_m}^{(N,M)}; \\[0.2cm] 
(M,n_m-N)\in \left( 2\mathbb{N}+1\right) ^{2}:
\lambda_m<-\lambda_{n_m}^{(N,M)} \text{ or } \lambda_m>0.%
\end{array}
\right.  \label{eq:reg-m-step-bis}
\end{equation}

%
%
Consider next the case where $m$ is odd, which leads to 
\begin{equation}
\mathrm{sign}\left( \lambda _{m}\right) = \mathrm{sign}\left(
(-1)^{M+1}\left( \left( -1\right) ^{n_{m}-N}\lambda _{m}-\lambda
_{n_{m}}^{\left( N,M\right) }\right) \right) .
\end{equation}
We then obtain the following possibilities: 
\begin{equation}
\left\{ 
\begin{array}{c}
(M,n_m-N)\in \left( 2\mathbb{N}\right) ^{2}:
0<\lambda_m<\lambda_{n_m}^{(N,M)}; \\[0.2cm] 
(M,n_m-N)\in 2\mathbb{N}\times \left( 2\mathbb{N}+1\right) :
\lambda_m<-\lambda_{n_m}^{(N,M)} \text{ or } \lambda_m>0; \\[0.2cm] 
(M,n_m-N)\in \left( 2\mathbb{N}+1\right) \times 2\mathbb{N}:
\lambda_m>\lambda_{n_m}^{(N,M)} \text { or } \lambda_m<0; \\[0.2cm] 
(M,n_m-N)\in \left( 2\mathbb{N}+1\right) ^{2}: -
\lambda_{n_m}^{(N,M)}<\lambda_m<0.%
\end{array}
\right.  \label{eq:reg-m-step-ter}
\end{equation}

%
%
Provided conditions (\ref{eq:reg-m-step-bis}) or (\ref{eq:reg-m-step-ter})
are satisfied, the result will still be valid at the $m$th step.

%
%

\subsection{Rationally-extended potential}

If $\lambda _{1},\lambda _{2},\ldots ,\lambda _{m}$ satisfy the
above-mentioned conditions, the $m$-step chain generates a regular potential 
\begin{align}
& V^{(N_{m})}(x;N,M;\lambda _{1},\ldots ,\lambda _{m})  \notag \\
& \quad =V(x;N,M)-2\frac{d^{2}}{dx^{2}}\log W(\psi _{-n_{1}-1}(x;N,M;\lambda
_{1}),\ldots ,\psi _{-n_{m}-1}(x;N,M;\lambda _{m})\mid x),
\end{align}
which turns out to be a rational extension of $V(x;N-m,M-m)$, 
\begin{align}
& V^{(N_{m})}(x;N,M;\lambda _{1},\ldots ,\lambda _{m})  \notag \\
& \quad =V(x;N-m,M-m)+E_{-m}(N,M)  \notag \\
& \qquad -8(1-z^{2})\frac{d^{2}}{dz^{2}}\log \left( \det \left[ \mathbf{R}%
\left( z;\lambda _{1},...,\lambda _{m}\right) \right] \right)  \notag \\
& \qquad +8z\frac{d}{dz}\log \left( \det \left[ \mathbf{R}(z;\lambda
_{1},...,\lambda _{m})\right] \right) ,  \label{eq:V-ext-m}
\end{align}
where the matrix elements of $\mathbf{R}(z;\lambda _{1},...,\lambda _{m})$ are given in Eq.~(\ref{mat
R 2}).
\par
%
The derivation of (\ref{eq:V-ext-m}) proceeds as that of (\ref{eq:V-ext}) in
Sec.~V by using the identities 
\begin{align}
& W(\psi _{-n_{1}-1}(x;N,M;\lambda _{1}),\psi _{-n_{2}-1}(x;N,M;\lambda
_{2}),\ldots ,\psi _{-n_{m}-1}(x;N,M;\lambda _{m})\mid x)  \notag \\
& \quad =\psi _{-1}^{m}(x;N,M)\left( \frac{dz}{dx}\right) ^{m(m-1)/2}\det %
\left[ \mathbf{R}\left( z;\lambda _{1},...,\lambda _{m}\right) \right] ,
\end{align}
\begin{equation}
\psi _{-1}^{m}(x;N,M)\left( \frac{dz}{dx}\right) ^{m(m-1)/2}\propto
\prod_{i=0}^{m-1}\psi _{-1}(x;N-i,M-i),
\end{equation}
\begin{equation}
\sum_{i=0}^{m-1}E_{-1}(N-i,M-i)=E_{-m}(N,M),
\end{equation}
as well as the shape invariance property (\ref{eq:shape}) of the TDPT
potential.
\par
%
%
{}For $V^{\left( N_{m}\right) }\left( x;N,M;\lambda _{1},\ldots,\lambda
_{m}\right) $ to be a confining potential on $]0,\pi /2[$, we need to impose
that $V\left( x;N-m,M-m\right) $ has this property, which is achieved for $%
N,M>m+1/2$, thence $N,M\geq m+1$. For such $N,M$ values, we note that $%
V^{\left( N_{m}\right) }\left( x;N,M;\lambda _{1},\ldots,\lambda _{m}\right) 
$ is actually strongly repulsive in both $0$ and $\pi /2$, since the
singularities are there of the type $g/x^{2}$ ($g\geq 3/4$) and $g/\left(
\pi /2-x\right) ^{2}$ ($g\geq 3/4$), respectively. This means that at each
extremity, only one basis solution is quadratically integrable \cite{frank}.
\par
%
%
The corresponding eigenstates are given by
\begin{align}
  &\psi^{(N_m)}_k(x;N,M;\lambda_1,\ldots,\lambda_m)  \notag \\
  &\quad = \frac{W(\psi_{-n_1-1}(x;N,M;\lambda_1), \ldots,
      \psi_{-n_m-1}(x;N,M;\lambda_m), \psi_k(x;N,M) \mid x)}{W(\psi_{-n_1-1}(x;N,M;%
      \lambda_1), \ldots, \psi_{-n_m-1}(x;N,M;\lambda_m \mid x)},
      \label{eq:psi-1-m}
\end{align}
for $k=0,1,2,\dots$, and 
\begin{align}
  &\psi^{(N_m)}_{-n_i-1}(x;N,M;\lambda_1, \ldots, \lambda_m)  \notag \\
  &\quad = \frac{W\left(\psi_{-n_1-1}(x;N,M;\lambda_1), \ldots, \check{\psi}%
     _{-n_i-1}(x;N,M;\lambda_i), \ldots,\psi_{-n_m-1}(x;N,M;\lambda_m) \mid
     x\right)}{W(\psi_{-n_1-1}(x;N,M;\lambda_1), \ldots,
     \psi_{-n_m-1}(x;N,M;\lambda_m) \mid x)},  \label{eq:psi-2-m}
\end{align}
for $i=1$, 2, \ldots, $m$, with corresponding energies $E_k(N,M)$, $k=0$, 1,
2, \ldots\ and $E_{-n_i-1}$, $i=1$, 2, \ldots, $m$, respectively. Here, $%
\check{\psi}_{-n_i-1}(x;N,M;\lambda_i)$ means that $\psi_{-n_i-1}(x;N,M;%
\lambda_i)$ is excluded from the Wronskian.
\par
%
%
Eq.~(\ref{deriv Jac 1}) can be generalized as \cite{GGM}
\begin{equation}
  \frac{d^{l}}{dz^{l}}\left( (1-z)^{N}(1+z)^{M}P_{k}^{(N,M)}(z)\right) =\left(
  -2\right) ^{l}(k+l)_{\underline{l}%
  }(1-z)^{N-l}(1+z)^{M-l}P_{k+l}^{(N-l,M-l)}(z).  \label{deriv Jac 2}
\end{equation}
\par
%
%
Using (\ref{wronsk prop}), (\ref{mat R}), and (\ref{deriv Jac 2}), Eq.~(\ref{eq:psi-1-m}) can be rewritten as
\begin{align}
  &\psi _{k}^{(N_{m})}(x;N,M;\lambda _{1},\ldots ,\lambda _{m}) \nonumber \\
  &\quad =\psi_{-1}(x;N,M)\left( \frac{dz}{dx}\right) ^{m} \nonumber\\
  &\qquad\times \frac{W(p_{n_{1}}^{(-N,-M)}(z;\lambda _{1}),\ldots
    ,p_{n_{m}}^{(-N,-M)}(z;\lambda _{m}),(1-z)^{N}(1+z)^{M}P_{k}^{(N,M)}(z)\mid
    z)}{W\left( p_{n_{1}}^{(-N,-M)}(z;\lambda
    _{1}),p_{n_{2}}^{(-N,-M)}(z;\lambda _{2}),\ldots
    ,p_{n_{m}}^{(-N,-M)}(z;\lambda _{m})\mid z\right) } \nonumber\\
  &\quad= W(p_{n_{1}}^{(-N,-M)}(z;\lambda _{1}),\ldots
    ,p_{n_{m}}^{(-N,-M)}(z;\lambda _{m}),(1-z)^{N}(1+z)^{M}P_{k}^{(N,M)}(z)\mid z) \nonumber\\
  &\qquad\times \frac{\psi _{-1}(x;N-m,M-m)}{\det \left[ \mathbf{R}\left( z;\lambda
    _{1},...,\lambda _{m}\right) \right] },
\end{align}
where the last Wronskian can be written as
\begin{align}
  &W(p_{n_{1}}^{(-N,-M)}(z;\lambda _{1}),\ldots
    ,p_{n_{m}}^{(-N,-M)}(z;\lambda _{m}),(1-z)^{N}(1+z)^{M}P_{k}^{(N,M)}(z)\mid z) \nonumber\\
  &\quad =(1-z)^{N-m}(1+z)^{M-m}Q_{k}^{(n_{1},n_{2},\ldots ,n_{m})}(z;N,M;\lambda
    _{1},\ldots ,\lambda _{m})
\end{align}
with
\begin{eqnarray*}
  &&Q_{k}^{(n_{1},n_{2},\ldots ,n_{m})}(z;N,M;\lambda _{1},\ldots ,\lambda_{m}) \\
&=&\left\vert 
\begin{array}{cccc}
\mathbf{R}_{1,1}\left( z;\lambda _{1},...,\lambda _{m}\right) & ... & 
\mathbf{R}_{1,m}\left( z;\lambda _{1},...,\lambda _{m}\right) & 
(1-z^{2})^{m}P_{k}^{(N,M)}(z) \\ 
... &  & ... & ... \\ 
\mathbf{R}_{m,1}\left( z;\lambda _{1},...,\lambda _{m}\right) & ... & 
\mathbf{R}_{m,m}\left( z;\lambda _{1},...,\lambda _{m}\right) & \left(
-2\right) ^{m-1}(k+m-1)_{\underline{m-1}}\\
& & & \times(1-z^{2})P_{k+m-1}^{(N-m+1,M-m+1)}(z)\\ 
\mathbf{R}_{m+1,1}\left( z;\lambda _{1},...,\lambda _{m}\right) & ... & \mathbf{%
R}_{m+1,m}\left( z;\lambda _{1},...,\lambda _{m}\right) & \left( -2\right)
^{m}(k+m)_{\underline{m}}P_{k+m}^{(N-m,M-m)}(z)%
\end{array}%
\right\vert .
\end{eqnarray*}
\par
%
%
We then get
\begin{align}
  & \psi _{k}^{(N_{m})}(x;N,M;\lambda _{1},\ldots ,\lambda _{m})  \notag \\
  & \quad =\frac{\psi _{-1}(x;N-m,M-m)(1-z)^{N-m}(1+z)^{M-m}}{\det \left[ 
    \mathbf{R}\left( z;\lambda _{1},...,\lambda _{m}\right) \right] }\times
    Q_{k}^{(n_{1},n_{2},\ldots ,n_{m})}(z;N,M;\lambda _{1},\ldots ,\lambda _{m})\notag \\
  & \quad =\frac{\psi _{0}(x;N-m,M-m)}{\det \left[ \mathbf{R}\left( z;\lambda
    _{1},...,\lambda _{m}\right) \right] }\times Q_{k}^{(n_{1},n_{2},\ldots
    ,n_{m})}(z;N,M;\lambda _{1},\ldots ,\lambda _{m}),
\end{align}
for $k=0,1,2,\ldots $.
\par
%
%
Moreover,
\begin{align}
  & \psi _{-n_{i}-1}^{(N_{m})}(x;N,M;\lambda _{1},\ldots ,\lambda _{m}) \notag \\
  &=\frac{1}{\psi _{-1}(x;N,M)\left( \frac{dz}{dx}\right) ^{m-1}}  \notag \\
  & \times \frac{W(p_{n_{1}}^{(-N,-M)}(z;\lambda
    _{1}),\ldots,\check{p}_{n_i}^{(-N,-M)}(z;\lambda_i),\ldots
    ,p_{n_{m}}^{(-N,-M)}(z;\lambda _{m})\mid z)}{W\left(
    p_{n_{1}}^{(-N,-M)}(z;\lambda _{1}),\ldots ,p_{n_{m}}^{(-N,-M)}(z;\lambda
    _{m})\mid z\right) },
\end{align}
where $\check{p}_{n_i}^{(-N,-M)}(z;\lambda_i)$ means that $p_{n_i}^{(-N,-M)}(z;\lambda_i)$ is excluded from the Wronskian. This leads to
\begin{align}
  &\psi^{(N_m)}_{-n_i-1}(x;N,M;\lambda_1, \ldots, \lambda_m) \nonumber \\
  &\quad = \frac{\psi_0(x;N-m,M-m)}{\det[\mathbf{R}(z;\lambda_1, \ldots, \lambda_m)]} 
     \times Q^{(n_1,n_2, \ldots, n_m)}_{-n_i-1}(z;N,M;\lambda_1, \ldots, \lambda_m),
\end{align}
where $Q^{(n_1,n_2;\ldots,n_m)}_{-n_i-1}(x;N,M;\lambda_1,\ldots,\lambda_m)$ is obtained from the determinant of $\mathbf{R}(z;\lambda_1, \ldots, \lambda_m)$ by suppressing the $m^{\rm th}$ row and the $i^{\rm th}$ column, i.e., by a minor that we denote by $\det[\mathbf{R}(z;\lambda_1, \ldots, \lambda_m)]^{m,i}$,
\begin{equation}
  Q^{(n_1,n_2;\ldots,n_m)}_{-n_i-1}(x;N,M;\lambda_1,\ldots,\lambda_m) = 
  \det[\mathbf{R}(z;\lambda_1, \ldots, \lambda_m)]^{m,i}.
\end{equation}
\par
%
%
Due to the orthogonality properties of the $\psi _{k}^{(N_{m})}$, the
polynomials $Q_{k}^{(n_{1},n_{2},\ldots ,n_{m})}(z;N,M;\lambda _{1},\ldots
,\lambda _{m})$, $k=-n_{1}-1$, \ldots , $-n_{m}-1,0,1,2,\ldots $, form a set
of orthogonal polynomials on $]-1,1[$ with respect to the measure 
\begin{align}
  & \mu _{n_{1}n_{2}\ldots n_{m}}^{(-N,-M)}(z;\lambda _{1},\lambda
    _{2},\ldots ,\lambda _{m})  \notag \\
  & \quad =\frac{(1-z)^{N-m}(1+z)^{M-m}}{\left( \det \left[ \mathbf{R}\left(
    z;\lambda _{1},...,\lambda _{m}\right) \right] \right) ^{2}}.
\end{align}
\par
%
%
\section{CONCLUSION}

In this article, we have studied the multi-step version of a previous
construction of the TDPT potential regular rational extensions obtained by
one-step DT using seed functions associated with the para-Jacobi polynomials
of Calogero and Yi. We have shown that the eigenstates of such $m$-step
extensions are expressed in terms of novel families of EOPs, orthogonal on $%
]-1,+1[$ and depending not only on $m$ discrete parameters, but also on $m$
real continuous parameters $\lambda_1$, $\lambda_2$, \ldots, $\lambda_m$.
The sets of parameters are related by some restrictions coming from the TDPT
rational extensions regularity conditions, which we study in detail.
\par
%
%
Looking for possible relationships between our results and those obtained by
using other approaches for generating EOPs depending on an arbitrary number
of continuous parameters would be an interesting open question for future
work.
\par
%
%
\section*{ACKNOWLEDGMENTS}

CQ was supported by the Fonds de la Recherche Scientifique-FNRS under Grant
No.~4.45.10.08.
\par
%
%
\section*{AUTHOR DECLARATIONS}

\subsection*{Conflict of interest}

The authors have no conflict to disclose.
\par
%
%
\subsection*{Author Contributions}

\noindent Y.\ Grandati: Conceptualization (equal); Formal analysis (equal);
Investigation (equal); Methodology (equal); Writing - original draft
(equal); Writing - review \& editing (equal).

\noindent C.\ Quesne: Conceptualization (equal); Formal analysis (equal);
Investigation (equal); Methodology (equal); Writing - original draft
(equal); Writing - review \& editing (equal).
\par
%
%
\section*{DATA AVAILABILITY}

Data sharing is not applicable to this article since no new data were
created or analyzed in this study.
\par
%
%


\begin{thebibliography}{99}

\bibitem{gomez09} D.\ G\'omez-Ullate, N.\ Kamran, and R.\ Milson, ``An
extended class of orthogonal polynomials defined by a Sturm-Liouville
problem,'' J.\ Math.\ Anal.\ Appl.\ \textbf{359}, 352 (2009);
arXiv:0807.3939.

\bibitem{gomez13a} D.\ G\'omez-Ullate, F.\ Marcell\'an, and R.\ Milson,
``Asymptotic and interlacing properties of zeros of exceptional Jacobi and
Laguerre polynomials,'' J.\ Math.\ Anal.\ Appl.\ \textbf{399}, 480 (2013);
arXiv:1204.2282.

\bibitem{kuijlaars} A.\ B.\ J.\ Kuijlaars and R.\ Milson, ``Zeros of
exceptional Hermite polynomials,'' J.\ Approximation Theory \textbf{200}, 28
(2015); arXiv:1412.6364.

\bibitem{duran15} A.\ J.\ Dur\'an, ``Higher order recurrence relation for
exceptional Charlier, Meixner, Hermite and Laguerre orthogonal
polynomials,'' Integral Transforms Spec.\ Funct.\ \textbf{26}, 357 (2015);
arXiv:1409.4697.

\bibitem{miki} H.\ Miki and S.\ Tsujimoto, ``A new recurrence formula for
generic exceptional orthogonal polynomials,'' J.\ Math.\ Phys.\ \textbf{56},
033502 (2015); arXiv:1410.0183.

\bibitem{odake16} S.\ Odake, ``Recurrence relations of the multi-indexed
orthogonal polynomials.\ III,'' J.\ Math.\ Phys.\ \textbf{57}, 023514
(2016); arXiv:1509.08213.

\bibitem{gomez13b} D.\ G\'omez-Ullate, N.\ Kamran, and R. Milson, ``A
conjecture on exceptional orthogonal polynomials,'' Found.\ Comput.\ Math.\ 
\textbf{13}, 615 (2013); arXiv:1203.6857.

\bibitem{garcia19} M.\ A.\ Garc\'\i a-Ferrero, D.\ G\'omez-Ullate, and R.\
Milson, ``A Bochner type characterization theorem for exceptional orthogonal
polynomials,'' J.\ Math.\ Anal.\ Appl.\ \textbf{472}, 584 (2019);
arXiv:1603.04358.

\bibitem{cq08} C.\ Quesne, ``Exceptional orthogonal polynomials, exactly
solvable potentials and supersymmetry,'' J.\ Phys.\ A:\ Math.\ Theor.\ 
\textbf{41}, 392001 (2008); arXiv:0807.4087.

\bibitem{cq09} C.\ Quesne, ``Solvable rational potentials and exceptional
orthogonal polynomials in supersymmetric quantum mechanics,'' SIGMA \textbf{5%
}, 084 (2009); arXiv:0906.2331.

\bibitem{odake09} S.\ Odake and R.\ Sasaki, ``Infinitely many shape
invariant potentials and new orthogonal polynomials,'' Phys.\ Lett.\ B 
\textbf{679}, 414 (2009); arXiv:0906.0142.

\bibitem{gomez12} D.\ G\'{o}mez-Ullate, N.\ Kamran, and R.\ Milson,
\textquotedblleft Two-step Darboux transformations and exceptional Laguerre
polynomials,\textquotedblright\ J.\ Math.\ Anal.\ Appl.\ \textbf{387}, 410
(2012); arXiv:1103.5724.

\bibitem{odake11} S.\ Odake and R.\ Sasaki, ``Exactly solvable quantum
mechanics and infinite families of multi-indexed orthogonal polynomials,''
Phys.\ Lett.\ B \textbf{702}, 164 (2011); arXiv:1105.0508.

\bibitem{cq11} C.\ Quesne, ``Rationally-extended radial oscillators and
Laguerre exceptional orthogonal polynomials in $k$th-order SUSYQM,'' Int.\
J.\ Mod.\ Phys.\ A \textbf{26}, 5337 (2011); arXiv:1110.3958.

\bibitem{grandati12} Y.\ Grandati, ``Multistep DBT and regular rational
extensions of the isotonic oscillator,'' Ann.\ Phys.\ (N.\ Y.) \textbf{327},
2411 (2012); arXiv:1108.4503.

\bibitem{schulze} A.\ Schulze-Halberg and B.\ Roy, ``Darboux partners of
pseudoscalar Dirac potentials associated with exceptional orthogonal
polynomials,'' Ann.\ Phys.\ (N.\ Y.) \textbf{349}, 159 (2014);
arXiv:1409.0999.

\bibitem{marquette} I.\ Marquette and C.\ Quesne, ``New families of
superintegrable systems from Hermite and Laguerre exceptional orthogonal
polynomials,'' J.\ Math.\ Phys.\ \textbf{54}, 042102 (2013); arXiv:1211.2957.

\bibitem{post} S.\ Post, S.\ Tsujimoto, and L.\ Vinet, ``Families of
superintegrable Hamiltonians constructed from exceptional polynomials,'' J.\
Phys.\ A: Math.\ Theor.\ \textbf{45}, 405202 (2012); arXiv:1206.0480.

\bibitem{clarkson} P.\ A.\ Clarkson, D.\ G\'omez-Ullate, Y.\ Grandati, and
R.\ Milson, ``Cyclic Maya diagrams and rational solutions of higher order
Painlev\'e systems,'' Stud.\ Appl.\ Math.\ \textbf{144}, 357 (2020);
arXiv:1811.09274.

\bibitem{bonneux} N.\ Bonneux, ``Exceptional Jacobi polynomials,'' J.\
Approximation Theory \textbf{239}, 72 (2019); arXiv:1804.01323.

\bibitem{duran17} A.\ J.\ Dur\'an, ``Exceptional Hahn and Jacobi orthogonal
polynomials,'' J.\ Approximation Theory \textbf{214}, 9 (2017);
arXiv:1510.02579.

\bibitem{szego} G.\ Szeg\"{o}, \textit{Orthogonal Polynomials} (American
Mathematical Society, Providence, RI, 1975).

\bibitem{calogero} F.\ Calogero and G.\ Yi, "Can the \textit{general}
solution of the second-order ODE characterizing Jacobi polynomials be 
\textit{polynomial}?", J.\ Phys.\ A: Math.\ Theor. \ \textbf{45}, 095206
(2012).

\bibitem{bagchi} B. Bagchi, Y. Grandati and C. Quesne, "Rational extensions
of the trigonometric Darboux-P\"{o}schl-Teller potential based on
para-Jacobi polynomials", J.\ Math.\ Phys.\ \textbf{56}, 062103 (2015);
arXiv:1411.7857.

\bibitem{grandati15} Y.\ Grandati and C.\ Quesne, ``Confluent chains of DBT:
enlarged shape invariance and new orthogonal polynomials,'' SIGMA \textbf{11}%
, 061 (2015); arXiv:1503.07747.

\bibitem{garcia21} M.\ A.\ Garc\'\i a-Ferrero, D.\ G\'omez-Ullate, and R.\
Milson, ``Exceptional Legendre polynomials and confluent Darboux
transformations,'' SIGMA \textbf{17}, 016 (2021); arXiv:2008.02822.

\bibitem{garcia22} M.\ A.\ Garc\'\i a-Ferrero, D.\ G\'omez-Ullate, R.\
Milson, and J.\ Munday, ``Exceptional Gegenbauer polynomials via isospectral
deformations,'' Stud.\ Appl.\ Math.\ \textbf{149}, 324 (2022);
arXiv:2110.04059.

\bibitem{duran21} A.\ J.\ Dur\'an, ``Exceptional Hahn and Jacobi polynomials
with an arbitrary number of continuous parameters,'' Stud.\ Appl.\ Math.\ 
\textbf{148}, 606 (2021).

\bibitem{duran23} A.\ J.\ Dur\'an, ``Exceptional Jacobi polynomials which
are deformations of Jacobi polynomials,'' J.\ Math.\ Anal.\ Appl.\ \textbf{%
528}, 127523 (2023).

\bibitem{crum} M.\ M.\ Crum, ``Associated Sturm-Liouville systems,'' Q.\ J.\
Math.\ Oxford Ser.\ 2 \textbf{6}, 121 (1955).

\bibitem{muir} T.\ Muir, \textit{A Treatise on the Theory of Determinants}
(Dover, NewYork, 1960), revised and enlarged by W.\ H.\ Metzler.

\bibitem{footnote} These inequalities correct Eq.~(4.7) of Ref.~\cite{bagchi}.

\bibitem{frank} W.\ M.\ Frank, D.\ J.\ Land, and R.\ M.\ Spector, "Singular
potentials", Rev.\ Mod.\ Phys.\ \textbf{43}, 36 (1971).

\bibitem{koekoek} R.\ Koekoek, P.\ A.\ Lesky, and R.\ F.\ Swarttouw, \textit{Hypergeometric Orthogonal Polynomials and Their $q$-Analogues} (Springer, Heidelberg, 2010).

\bibitem{GGM} D.\ G\'{o}mez-Ullate, Y.\ Grandati and R.\ Milson,
\textquotedblleft Shape invariance and equivalence relations for
pseudo-Wronskians of Laguerre and Jacobi polynomials,\textquotedblright\ J.\
Phys.\ A: Math.\ Theor.\ \textbf{51}, 345201 (2018).

\end{thebibliography}
\end{document}